\documentclass{elsart}
\usepackage{epsfig}
\usepackage[scanall]{psfrag}
\usepackage{supertabular}
\usepackage{amsmath}
\def\rvec{{\bf r}}				
\def\dens{\rho}					
\def\labelssize{\footnotesize}            
\def\titlessize{\scriptsize}              
\def\legendsize{\scriptsize}	            
\def\skala{1.0}                           
\begin{document}
\begin{frontmatter}
\title{{\em Ab initio} pseudopotentials for electronic
structure calculations of poly-atomic systems using
density-functional theory.}
\author{Martin Fuchs\!\thanksref{email}}\author{\space and Matthias Scheffler}
\address{Fritz-Haber-Institut der Max-Planck-Gesellschaft, Faradayweg 4-6,
D-14195 Berlin-Dahlem, Germany}
\thanks[email]{E-Mail: fuchs@fhi-berlin.mpg.de}
\begin{abstract}
The package {\sf fhi98PP} allows one to generate norm-conserving
pseudopotentials adapted to density-functional theory total-energy
calculations for a multitude of elements throughout the periodic 
table, including first-row and transition metal elements. 
The package also facilitates a first assessment of the 
pseudopotentials' transferability, either in semilocal or fully 
separable form, by means of simple tests carried out for the free 
atom. Various parameterizations of the local-density approximation 
and the generalized gradient approximation for exchange and 
correlation are implemented.
\end{abstract}
\end{frontmatter}

\vskip 4em
\clearpage

\twocolumn
\section*{PROGRAM SUMMARY}

\sloppy
{\em Title of the program:} {\sf fhi98PP}\\[1ex]
{\em catalogue number:} ... \\[1ex]
{\em Program obtainable from:}
 CPC Program Library, Queen's University of Belfast, N. Ireland
(see application form in this issue)\\[1ex]
{\em Licensing provisions:} none\\[1ex]
{\em Computers on which the program has been developed:}
IBM/\,RS 6000, Pentium PC\\[1ex]
{\em Operating system:} UNIX\\[1ex]
{\em Graphics software to which output is tailored:}
XMGR or XVGR (both are public domain packages)\\[1ex]
{\it Programming language:} FORTRAN~77\\
(non-standard feature is the use of \verb|END DO|);
UNIX C-shell scripts are employed as command line 
interfaces\\[1ex]
{\em Memory required to execute with typical data:}
$<$ 1~Mbyte\\[1ex]
{\em No.~of bits in a word:} 32\\[1ex]
{\em Memory required for test run:}  $<$ 1~Mbyte\\[1ex]
{\em Time for test run:} $\approx$ 1\,min\\[1ex]
{\em Number of lines in distributed program:} 7500 
(19000 with linear algebra library)\\[1ex]
{\em PACS codes:} 71.15.Hx, 71.15.Mb, 71.15.-m, 82.20.Wt\\[1ex]
{\em Keywords:} 
pseudopotential, total energy, electronic structure, density functional, 
local density, generalized gradient

{\em Nature of the physical problem}\\[1ex]
The norm-conserving pseudopotential concept allows for efficient
and accurate {\em ab initio} electronic structure calculations of
poly-atomic systems. The key features of this approach are:
(i) Only the valence states need to be calculated. The core
states are considered as chemically inert, and removed within 
the frozen-core approximation. This exploits that many chemical 
and physical processes are governed by the valence states but 
only indirectly involve the core states.
(ii) The valence electrons move in a pseudopotential which is much
smoother than the true potential inside the small core 
regions around the nuclei, while reproducing it outside. This 
pseudopotential acts on smooth pseudo wavefunctions that are equivalent 
to the true valence wavefunctions, but avoid the radial nodes that keep 
the true valence and core orbitals orthogonal. This enables
the use of computationally expedient basis sets like plane waves,
and facilitates the numerical solution of the Schr\"odinger and 
Poisson equations in complicated systems.
(iii) The norm-conservation constraint ensures that outside the core
the pseudo wavefunctions behave like their all-electron 
counterparts over a wide range of different chemical situations. 
Along with a proper design, this makes the pseudopotential approach 
a dependable approximation in describing chemical bonds.

Derived and applied within density-functional 
theory~\cite{hoh64a,dre91a,koh96a},
norm-conserving pseudopotentials~\cite{top73a,ham79a,bac82a} enable 
total-energy calculations of complex poly-atomic systems
~\cite{pic89a,pay92a,boc97a} for a multitude of elements throughout 
the periodic table. Questions addressed with pseudopotentials provided 
by this code, or its earlier version, range from phase 
transitions~\cite{mol95a,eng97a}, defects in 
semiconductors~\cite{boc97b,sch98a,sta98a}, the structure of and diffusion on 
surfaces of semiconductors~\cite{mol96a,kle97a,lot98a}, 
simple metals~\cite{stu97a}, 
and transition metals~\cite{yu96a,rat97a,boi97a}, up to surface 
reactions~\cite{peh95a,gro97a}, including molecules~\cite{sch96a,sta96b} 
of first-row species.

This package is a tool to generate and validate norm-conserving 
pseudopotentials, usable either in semilocal or in fully separable form,
and including relativistic effects. Exchange and correlation is treated 
in the local-density approximation based on Ceperley and Alder's 
data~\cite{cep80a} as parametrized, e.g., by Perdew and Wang~\cite{per92b}, 
or in the generalized gradient approximation, as proposed by Perdew, Burke, 
and Ernzerhof (PBE)~\cite{per96b}, 
Perdew and Wang (PW91)~\cite{per92a}, 
Becke and Perdew (BP)~\cite{bec88a,per86a}, 
and by Lee, Yang, and Parr (BLYP)~\cite{lee88a}. 

{\em Method of solution}\\[1ex]
The first part of the program \mbox{({\sf psgen})} generates 
pseudopotentials of the Hamann~\cite{ham89a} or the 
Troullier-Martins type~\cite{tro91a}, based on a 
scalar-relativistic all-electron calculation of the 
free atom. 
A partial core density can be included to allow for nonlinear core-valence 
exchange-correlation~\cite{lou82a} where needed, e.g., for spin-density
functional calculations, alkali metal compounds, and the cations of II-VI 
compounds like ZnSe.
The second part ({\sf pswatch}) serves to assess the transferability of the pseudopotentials,
examining scattering properties, excitation energies, and chemical hardness 
properties of the free pseudo atom. Transcribing the pseudopotentials into
the fully separable form of Kleinman and Bylander~\cite{kle82a}, we verify
the absence of unphysical states by inspection of the bound state spectrum 
and by the analysis of Gonze {\it et al.}~\cite{gon91a}.
The convergence of the pseudo wavefunctions in momentum space is monitored 
in order to estimate a suitable basis set cutoff in plane wave calculations.

{\em Restrictions on the complexity of the problem}\\[1ex]
(i) Only some of the GGA's currently in use are implemented, 
others may be readily added however. (ii) The present 
pseudopotentials yield the correct relativistic 
valence levels where spin-orbit splittings are 
averaged over, as it is intended for most applications. 

\onecolumn

\section{Introduction}
\label{sec:introduction}

Electronic structure and total-energy calculations using
density-functional theory~\cite{hoh64a,dre91a,koh96a} have 
grown into a powerful theoretical tool to gain a quantitative 
understanding of the physics and chemistry of complex molecular, 
liquid, and solid state systems. The norm-conserving pseudopotential
approach provides an effective and reliable means for performing 
such calculations in a wide variety of poly-atomic systems,
particularly, though not only, together with a plane wave basis 
and modern minimization algorithms for the variational 
determination of the ground state~\cite{pic89a,pay92a,boc97a,kre96a}.
In this approach only the chemically active valence electrons 
are dealt with explicitly. The chemically inert core electrons 
are eliminated within the frozen-core approximation~\cite{bar80a},
being considered together with the nuclei as rigid non-polarizeable 
ion cores. In turn, all electrostatic and quantum-mechanical interactions 
of the valence electrons with the ion cores (the nuclear Coulomb attraction 
screened by the core electrons, Pauli repulsion and exchange-correlation 
between core and valence electrons) are accounted for by angular
momentum dependent pseudopotentials. These reproduce the true potential 
and valence orbitals outside a chosen core region but remain much weaker inside. 
The valence electrons are described by smooth pseudo orbitals which play the same 
role as the true orbitals, but avoid the nodal structure near the nuclei 
that keeps core and valence states orthogonal in an all-electron framework. 
The respective Pauli repulsion largely cancels the attractive parts of the 
true potential in the core region, and is built into the therefore 
rather weak pseudopotential~\cite{coh70a}.
This ``pseudoization'' of the valence wavefunctions along with
the removal of the core states eminently facilitates a numerically
accurate solution of the Schr\"odinger and Poisson equations,
and enables the use of plane waves as an expedient basis set in 
electronic structure calculations. 
By virtue of the norm-conserving property~\cite{top73a,ham79a} and 
when constructed properly pseudopotentials present a rather marginal 
approximation, and indeed allow for an adequate description of the 
valence electrons over the entire chemically relevant range of systems, 
i.e. atoms, molecules, and solids.

Here we supply a two-piece package to generate norm-conserving 
pseudopotentials adapted to density-functional electronic structure
calculations, and usable in semilocal or fully separable form.
Generic input files for many elements are distributed 
along with the package.\\[1ex]

Part 1 (program {\sf psgen}) serves to generate the pseudopotentials 
using the schemes by Hamann~\cite{ham89a} or by Troullier and Martins~\cite{tro91a}.
This combination provides efficient pseudopotentials for ``canonical''
applications like group IV and III-V 
semiconductors~\cite{boc97a,mol95a,mol96a,kle97a,lot98a,peh95a,gro97a}, 
as well as for systems where first-row, transition or noble metal elements
are present~\cite{eng97a,yu96a,rat97a,boi97a,sch96a,sta96b}, or 
where ``semicore'' $d$-states must be treated as valence states, like
in GaN~\cite{neu98a} or InN~\cite{sta98a}. In such cases the strongly 
localized $2p$ and $3,4,5d$ valence states are readily handled 
with the Troullier-Martins scheme. Both schemes have been routinely 
employed and have an established track record with the electronic 
structure and {\it ab initio} molecular dynamics package 
{\sf fhi96md}~\cite{boc97a}. 
The pseudopotentials are derived within density-functional theory,
starting from a scalar-relativistic all-electron calculation of the 
free atom~\cite{koe77a}. Accordingly they yield the proper relativistic 
positions of the valence levels~\cite{com:rel} with the spin-orbit 
coupling being averaged over, in line with the practice in most applications.
The exchange-correlation interaction may be described within the 
local-density approximation (LDA)~\cite{koh65a,jon89a}, 
or the generalized gradient approximation (GGA)~\cite{bur96c}.
For the LDA, the package includes the expressions by Perdew and 
Wang~\cite{per92b}, and by Perdew and Zunger~\cite{per81a}, both
parameterizing Ceperley and Alder's results for the homogeneous 
electron gas~\cite{cep80a}. Earlier prescriptions by Hedin and 
Lundquist~\cite{hed71a}, and Wigner~\cite{wig34a} are supplied as well. 
For the GGA, the package includes the formulations 
by Perdew, Burke, and Ernzerhof (PBE)~\cite{per96b}, 
by Perdew and Wang (PW91)~\cite{per92a},
as well as those composed from the exchange functional
by Becke~\cite{bec88a} (or that of Ref.~\cite{per92a}) and the correlation
functional either by Perdew (BP)~\cite{per86a}, or by Lee, Yang, and Parr
(BLYP)~\cite{lee88a}. Alternative forms can be readily added to the code.
Hence the pseudopotentials, i.e. the electron-ion interaction, can be
generated consistently using the same exchange-correlation scheme 
as for calculating the poly-atomic system~\cite{fuc98a}.
A partial core density may be included in the unscreening of the
pseudopotentials to explicitly account for the nonlinearity of 
core-valence exchange-correlation of the above functionals~\cite{lou82a}.
This has proven necessary, e.g., for alkali metal elements~\cite{heb92a},
the cations in strongly polar compounds like ZnSe~\cite{kle94a}, or
calculations of spin-polarized systems~\cite{cho96a}.

Part 2 (program {\sf pswatch}) facilitates the assessment of the 
pseudopotentials' transferability, that is, their proper performance
in different chemical environments. This is done for the free (pseudo) 
atom by checking that its scattering properties, excitation energies and 
chemical hardness properties \cite{tet93a,fil95a} faithfully represent 
those of the corresponding all-electron atom. 
For this purpose, we evaluate the logarithmic derivatives of the 
valence wavefunctions, total energies and orbital eigenvalues for different 
valence configurations. For the commonly used fully separable Kleinman-Bylander
form of the pseudopotentials~\cite{kle82a}, it is important to rule out 
``ghost states'', i.e. unphysical bound states or resonances appearing 
in the valence spectrum. As an immediate 
check we evaluate the bound state spectrum of the fully separable
potentials, and apply the analysis by Gonze {\it et al.}~\cite{gon91a}. 
To assess the convergence behavior of the pseudopotentials
we monitor the pseudo atom's kinetic energy in momentum 
space~\cite{kre94a}, which also allows a first estimate of a
suitable basis cutoff for plane wave calculations.

In Chapter~\ref{cha:concept} of this Communication we briefly survey 
the conceptual aspects of constructing and validating pseudopotentials, 
more practical topics being addressed towards the end of a section.
Chapter~\ref{cha:package} is meant as a reference guide to our package.

\section{General background and formalism}
\label{cha:concept}

A pseudopotential is constructed to replace the atomic all-electron
potential such that core states are eliminated and the valence electrons
are described by nodeless pseudo wavefunctions. In doing so the principal 
objectives to consider are 
(i) the transferability of the pseudopotential, its ability
to accurately describe the valence electrons in different atomic,
molecular, and solid-state environments. In self-consistent
total energy calculations this means that the valence states 
have the proper energies and lead to a properly normalized
electron distribution which in turn yields proper electrostatic 
and exchange-correlation potentials, particularly outside the core 
region, i.e. where chemical bonds build up.  
(ii) their efficiency, that is to keep the computational workload
in applications as low as possible, allowing to compute wavefunctions
and electron densities with as few basis functions (being ``soft'')
and operations as possible. 
Modern norm-conserving pseudopotentials, implemented here in the
variants proposed by Hamann \cite{ham89a} and by Troullier-Martins \cite{tro91a},
allow for a controlled ``compromise'' with respect to these (conflicting)
tasks, through appropriate choices and tests at several stages of 
their actual construction. 

Norm-conserving pseudopotentials are derived from an atomic reference 
state (Sec.~\ref{sec:atomic} and~\ref{sec:screened}), requiring that 
the pseudo and the all-electron valence eigenstates have the same (reference)
energies and the same amplitude (and thus density) outside a chosen core 
cutoff radius $r^{c}$. Normalizing the pseudo orbitals then implies that they 
include the same amount of charge in the core region as their all-electron 
counterparts, being ``norm-conserving''.  
At the reference energies, the pseudo and all-electron logarithmic derivatives                         
(which play the role of boundary conditions on the wavefunctions inside 
and outside the core region) agree for $r \geq r^{c}$ by construction.
Norm-conservation then ensures that the pseudo and all-electron logarithmic 
derivatives agree also around each reference level, to first order in the 
energy, by virtue of the identity
\begin{equation}
\label{eq:norm1}
-\frac{1}{2} r^2 |\psi(\epsilon;\rvec)|^2 \frac{d}{d\epsilon} 
\left\{
 \frac{d}{dr} \ln \psi(\epsilon;\rvec)
\right\}\Bigg|_{\,^{\scriptstyle r = r^{\rm c}}_{\scriptstyle \epsilon = \epsilon'}}
=
\int\!\!\int_0^{r^c} |\psi(\epsilon';\rvec)|^2 r^2 dr\,d\Omega
\quad,
\end{equation}
where $\psi(\epsilon;\rvec)$ denotes the regular solution of the 
Schr\"odinger equation at some energy $\epsilon$ (not necessarily 
an eigenfunction) for either the all-electron potential or the 
pseudopotential.
A norm-conserving pseudopotential thus exhibits the same scattering 
properties (logarithmic derivatives) as the all-electron potential in 
the neighborhood of the atomic eigenvalues, typically it does so over the 
entire energy range of valence bands or molecular orbitals, which
is an important prerequisite and measure for the pseudopotential's 
proper performance.

Regarding transferability the quality of a pseudopotential depends 
on (a) correct scattering properties, (b) the actual choice of the core 
cutoff radius $r^{c}$, 
larger values leading to softer pseudopotentials (more rapidly convergent 
in say plane wave calculations) but tending to decrease their accuracy,
as the pseudo wavefunctions and the pseudopotential begin to deviate from their
true counterparts even at radii that are relevant to bonding,
(c) an adequate account of the 
nonlinear exchange-correlation interaction between core and valence 
electrons (Sec.~\ref{sec:unscreen}), which is often treated linearly
through the pseudopotential and thus approximated,
(d) if applied, the quality of the Kleinman-Bylander 
form of the pseudopotential, where accidental spurious states need to
be avoided (Sec.~\ref{sec:kb}), (e) the validity of the frozen-core
approximation, i.e. the inclusion of all relevant states as valence 
states which may include upper semicore states in addition to the outermost 
atomic shell, dependent on the actual chemical context, say, if these 
hybridize with valence states of neighboring atoms 
(Sec.~\ref{sec:unscreen}), and (f) the treatment of higher angular 
momentum components which are just implicitly considered in the 
pseudopotential's construction (Sec.~\ref{sec:use-local}). 
As outlined in Sec.~\ref{sec:transferability} 
the points (a) -- (e) can be controlled with respect to transferability 
by verifying that the pseudo atom reproduces the all-electron atom's 
scattering properties and excited states. The latter essentially probe 
the atom's response or hardness properties due to a change in the occupation
of the valence orbitals~\cite{tet93a}. Such tests for the free atom are less 
conclusive for (e) and (f) which, though mostly uncritical, should be kept in 
mind as potential sources of pseudopotential errors in poly-atomic systems. 
 
Usually norm-conserving pseudopotentials are set up and derived
in terms of an angular momentum dependent, ``semilocal'' 
operator,
\begin{equation}
\label{eq:semilocal-pseudopotential}
\langle \rvec | \hat V^{\rm ps} | \rvec' \rangle =
V^{\rm loc}(r)\delta(\rvec-\rvec') + 
            \sum_{l=0}^{l_{\rm max}}\sum_{m=-l}^{m=l} Y^{*}_{lm}(\Omega_{\rvec})
                  \delta V^{\rm ps}_{l}(r)\frac{\delta(r-r')}{r^2}
                        Y_{lm}(\Omega_{\rvec'}) 
\quad,
\end{equation}
written in terms of a local pseudopotential $V^{\rm loc}(r)$, and 
$l$-dependent components $\delta V^{\rm ps}_{l}(r) = V^{\rm ps}_{l}(r) - V^{\rm loc}(r)$
which are confined to the core region and eventually vanish beyond some 
$l_{\rm max}$ (see Sec. \ref{sec:use-local}). For reasons of computational 
efficiency the semilocal form is often transcribed into the fully nonlocal 
Kleinman-Bylander form~\cite{kle82a} (Sec.~\ref{sec:kb}). 
Relativistic effects on the electrons, important 
for heavier elements, basically originate just in the core region. Whence 
they may be incorporated in the pseudopotentials~\cite{kle80a,hem93a} which
we do here within a scalar-relativistic approximation (Sec.~\ref{sec:atomic}). 
Accordingly, such ``relativistic'' pseudopotentials are formally employed 
with a non-relativistic Schr\"odinger equation.

The next subsections discuss the construction of this package's 
pseudopotentials, which in short proceeds as follows,
\begin{itemize}
\item{
	Density-functional calculation of the all-electron atom in a 
      reference state (ground state), within a scalar-relativistic framework
      and a chosen approximation for exchange-correlation (Sec.~\ref{sec:atomic}).
}
\item{
	Construction of the pseudo valence orbitals and (screened) pseudopotential
      components, observing the norm-conserving constraints (Sec.~\ref{sec:screened}).
}
\item{
      Removal of the electrostatic and exchange-correlation components due to 
      the valence electrons from the screened pseudopotential, optionally taking
      nonlinear core-valence effects into account. This ``unscreening'' yields
      the ionic pseudopotential representing the electron-ion interaction in 
      poly-atomic systems (Sec.~\ref{sec:unscreen}).
}
\item{
      Assessment of the pseudopotential's transferability (Sec.~\ref{sec:transferability}),
      in case of a transformation to the fully separable Kleinman-Bylander 
      form the exclusion of unphysical ``ghost states'' in the valence spectrum (Sec.~\ref{sec:kb}).
     }
\end{itemize}
Hartree atomic units are used throughout ($e^2  =  \hbar  =  m  = 1$), 
lengths being given in units of the Bohr radius $1\,\mbox{bohr} = 0.52918\,\mbox{\AA}$, 
and energies in units of $1\,\mbox{hartree} = 27.2116\,\mbox{eV}$.

\subsection{Atomic all-electron calculation}
\label{sec:atomic}

The initial step in constructing our pseudopotentials is an
all-electron calculation of the free atom in a reference configuration, usually 
its neutral ground-state. Using density-functional theory to handle the many-electron
interactions, the total energy of $N$ electrons in an external potential 
$V^{\rm ext}(\rvec)$ (for an atom the nuclear $-Z/r$ potential) is written as a functional 
of the electron density $\dens$,
\begin{equation}
E^{\rm tot}[\dens] = T[\dens] + E^{\rm XC}[\dens] + E^{\rm H}[\dens] +
\int -\frac{Z}{r} \dens(\rvec) d^3r\quad,
\label{eq:etot-dft}
\end{equation}
where $T[\dens]$ is the non-interacting kinetic energy, $E^{\rm XC}[\dens]$ 
the exchange and correlation energy, and 
$E^{\rm H}[\dens] = \frac{1}{2}\int \frac{\dens(\rvec)\dens(\rvec')}{|\rvec-\rvec'|}
d^3r\,d^3r'$ the electrostatic or Hartree energy from the electron-electron 
repulsion. The ground-state corresponds to the minimum of $E^{\rm tot}[\dens]$, 
and is found from the self-consistent solution of the Kohn-Sham equations
\begin{equation}
\label{eq:kohn-sham}
\left[ -\frac{1}{2}\nabla^2 + V^{\rm eff}[\dens;\rvec] - \epsilon_i \right]
\psi_{i}(\rvec) = 0 \quad,
\end{equation}
\begin{equation}
V^{\rm eff}[\dens;\rvec] = \int \frac{\dens(\rvec)}{|\rvec-\rvec'|} d^3r 
+\frac{\delta E^{\rm XC}[\dens]}{\delta \dens(\rvec)} + V^{\rm ext}(\rvec) \quad,
\end{equation}
\begin{equation}
\dens(\rvec) = \sum_{i} f_{i} |\psi_{i}(\rvec)|^2\quad,\quad \int \dens(\rvec)d^3r =
N\quad,
\end{equation}
\begin{equation}
\label{eq:kohn-sham-f}
f_{i} = 1 \quad\mbox{if}\quad \epsilon_{i} \leq \epsilon_{N} \quad,\,\,
0 \leq f_{i} \leq 1 \quad\mbox{if}\quad \epsilon_{i} =  \epsilon_{N}\quad,\,\,
\mbox{ and}\quad f_{i} = 0 \quad\mbox{if}\quad \epsilon_{i} > \epsilon_{N}
\quad.
\end{equation}
Specializing to a spherically symmetric electron density 
and effective potential the Kohn-Sham orbitals separate, 
$\psi_{i}(\rvec) = [u_{n_i l_i}(\epsilon_{i}; r)/r] Y_{l_i m_i}(\Omega_{\rvec})$
and (\ref{eq:kohn-sham}) reduces to a radial Schr\"odinger 
equation. Relativistic effects, which in principle require 
a four-current and Dirac-spinor formulation of Eqs.~(\ref{eq:etot-dft}) -- 
(\ref{eq:kohn-sham-f}), are treated by using a scalar-relativistic kinetic 
energy operator~\cite{koe77a}. This takes into account the kinematic relativistic 
terms (the mass-velocity and Darwin terms) needed to properly describe 
the core states and the relativistic shifts of the valence 
levels, in particular for the heavier elements. The spin-orbit coupling 
terms are averaged over, as is commonly done in most applications~(see
also Ref.~\cite{hem93a}).
The radial wavefunctions $u_{nl}$ are then the self-consistent
eigenfunctions of the scalar-relativistic Schr\"odinger equation 
\begin{equation}
\label{eq:scalar-relativistic-ks}
\left[
\frac{1}{2M(r)}\left(
-\frac{d^{2}}{d\,r^{2}} 
- \frac{1}{2M(r)c^{2}} \frac{d\,V(r)}{d\,r} r \frac{d}{d\,r}\frac{1}{r}
+ \frac{l(l+1)}{r^{2}} 
\right)
+ V(r) - \epsilon_{i} \right]
u_{nl}(\epsilon_{i}; r)
= 0\,\,,
\end{equation}
with the relativistic electron mass $M(r) = 1 + \frac{\epsilon_{i} - V(r)}{2c^2}$, where
$1/c = 1/137.036$ is the finestructure constant, the effective all-electron potential
\begin{equation}
\label{eq:vae}
V(r) = V^{\rm XC}[\dens;r] + V^{\rm H}[\dens;r] -\frac{Z}{r}\quad,
\end{equation}
the exchange-correlation potential $V^{\rm XC}[\dens;r] = \frac{\delta E^{\rm XC}[\dens]}{\delta\dens(r)}$
\,, and the Hartree potential
\begin{equation}
\label{eq:vh}
V^{\rm H}[\dens;r] = \frac{\delta E^{\rm H}[\dens]}{\delta \dens(r)}
= 4\pi \frac{1}{r}\int_{0}^{r} \dens(r') {r'}^2 dr' +
4\pi \int_{r}^{\infty} \dens(r') r'dr' \quad.
\end{equation}
Due to the spherical symmetry, states with the same quantum numbers 
$n_{i}=n$ and $l_{i}=l$\,, but different $m_{i}$ are energetically 
degenerate, $\epsilon_i = \epsilon_{n_{i}l_{i}m_{i}} = \epsilon_{nl}$,
and therefore equally occupied by $0 \leq f_{i} \leq f_{nl}/(2l+1)$ 
electrons, where the occupancies $f_{nl}$ stand for the total number 
of electrons in the $nl$-shell. This leads to the electron density 
\begin{equation}
\label{eq:ground-density}
\dens(r)  =   \frac{1}{4\pi} \sum_{n=1}\sum_{l=0}^{n-1}
	f_{nl}\left|\frac{u_{nl}(\epsilon_{nl}; r)}{r}\right|^{2}\,\,,
	\mbox{with}\quad \left\{
\begin{array}{cl}
f_{nl}  =  2(2l+1) 	&\mbox{for}\quad \epsilon_{nl} < \epsilon_{N} \,,\\
0 \leq f_{nl} \leq 2(2l+1) \,&\mbox{for}\quad	
		\epsilon_{nl} = \epsilon_{N} \,,\\
f_{nl} = 0 	&\mbox{for}\quad  \epsilon_{nl} > \epsilon_{N}\,,
\end{array}\right.
\end{equation}
with the total number of electrons being $\sum_{nl} f_{nl} = N\,$. 
The non-relativistic Kohn-Sham equations are recovered by setting the finestructure 
constant $1/c$ in Eq.~(\ref{eq:scalar-relativistic-ks}) to zero. The above scheme 
determines the atomic ground state.
Excited atomic states are calculated with appropriately specified occupancies (as 
usual, we treat the occupancies always as parameters, a database of ground state 
configurations $\lbrace f_{nl}\rbrace$ being supplied with this package).

Regarding exchange-correlation, 
the program allows to employ commonly used 
parameterizations of the local-density approximation (LDA) and of 
the generalized gradient approximation (GGA). 
In the LDA the exchange-correlation energy is
\begin{equation}
\label{eq:exc-lda}
E^{\rm XC-LDA}[\dens]  =  \int \dens(r)\,\varepsilon^{\rm XC-LDA}(\dens(r)) d^3r\quad,
\end{equation}
where 
$\varepsilon^{\rm XC-LDA}(\dens)=
\varepsilon^{\rm X-LDA}(\dens)+\varepsilon^{\rm C-LDA}(\dens)$
represents the exchange-correlation energy per electron of a uniform electron gas. 
Its exchange part reads 
$\varepsilon^{\rm X-LDA}(\dens)=-\frac{3}{4\pi}(3\pi^2\dens)^{1/3}$. 
For the correlation part $\varepsilon^{\rm C-LDA}(\dens)$ representations given
by Perdew and Wang~\cite{per92a} and, previously, by Perdew and Zunger~\cite{per81a} 
may be used. Both are parameterizations of Ceperley and Alder's exact results for the 
uniform electron gas~\cite{cep80a}. To facilitate comparisons with results in the 
existing literature, we also supply the earlier prescriptions by Hedin and 
Lundquist~\cite{hed71a}, and by Wigner~\cite{wig34a}. For an overview of the LDA we 
refer to Ref.~\cite{jon89a}. These (nonrelativistic) LDA exchange-correlation 
functionals may be supplemented with relativistic corrections to the exchange part,
as given by MacDonald and Vosko~\cite{mac79a,com:relativity}.
The LDA exchange-correlation potential is given by
\begin{equation}
V^{\rm XC-LDA}[\dens;r]  =  \left[1 + \dens(r) \frac{d}{d\dens} \right]\varepsilon^{\rm XC-LDA}(\dens(r))
\quad.
\end{equation}

In a GGA the exchange-correlation functional depends on the density {\em and} 
its gradient,
\begin{equation} 
\label{eq:exc-gga}
E^{\rm XC-GGA}[\dens] = \int \dens(r) 
\varepsilon^{\rm XC-GGA}(\dens(r),|\nabla\dens(r)|)\, d^{3}r\quad.
\end{equation}
This code includes the GGA functionals by Perdew and Wang (PW91)~\cite{per92a}, by
Perdew, Burke, and Ernzerhof (PBE)~\cite{per96b}, and Becke's formula for the 
exchange part~\cite{bec88a} combined with Perdew's 1986 formula for 
correlation~\cite{per86a} (BP). Substituting for the latter the formula 
of Lee, Yang, and Parr (LYP)~\cite{lee88a} provides the so-called BLYP GGA. 
Relativistic corrections to the GGA as proposed by Engel {\em et al.}
~\cite{eng96a} are not included here. 
The GGA exchange-correlation potential (in cartesian coordinates) is given by
\begin{equation}
V^{\rm XC-GGA}[\dens;r] = \left[\frac{\partial}{\partial\dens} -
\sum_{i=1}^{3} \nabla_{i} \frac{\partial}{\partial\nabla_{i}\dens} 
\right]
\dens(r) \varepsilon^{\rm XC-GGA}(\dens(r),|\nabla\dens(r)|)\quad,
\end{equation}
and depends on the first and second radial derivatives of the density.
It is an open issue whether to prefer one of these GGAs over another.
Dependent on the application, they may yield somewhat differing results.
However, the actual choice among these GGAs is usually less important 
than the differences between the LDA and the GGAs themselves. A discussion 
of the various GGAs can be found in Ref.~\cite{bur96c}. 

Whether to use either the LDA or the GGA in constructing 
pseudopotentials is determined by the exchange-correlation scheme 
one wants to use for the poly-atomic system. If using the GGA 
there, the pseudopotential should be generated within the
(same) GGA rather than the LDA, in order to describe the 
exchange-correlation contribution to the core-valence 
interactions within the GGA too. This ``consistent'' use of the 
GGA may be significant to resolve any differences between LDA
and GGA properly, i.e. to the same extent as with all-electron 
methods~\cite{fuc98a} (see also Sec.~\ref{sec:unscreen}).

\subsection{Screened norm-conserving pseudopotentials}
\label{sec:screened}

Having obtained the all-electron potential and valence states, we use 
either the scheme by Hamann~\cite{ham89a}, or by Troullier and Martins~\cite{tro91a}
to construct the intermediate, so called screened pseudopotential. 
The latter acts as the effective potential on the (pseudo) valence states
$\psi^{\rm ps}_{lm}(\rvec)=\left[u^{\rm ps}_{l}(\epsilon^{\rm ps}_{l};r)/r \right]
Y_{lm}(\Omega_{\rvec})$, where the radial part for the angular momentum $l$ 
is the lowest eigenfunction of the nonrelativistic Schr\"odinger equation,
\begin{equation}
\label{eq:rseq}
\left[
-\frac{1}{2}\frac{d^2}{dr^2} + \frac{l(l+1)}{2r^2} + V^{\rm ps, scr}_{l}(r) 
			- \epsilon^{\rm ps}_{l}
\right]
u^{\rm ps}_{l}(\epsilon^{\rm ps}_{l};r) = 0
\quad.
\end{equation}
Initially, a radial pseudo wavefunction $u^{\rm ps}_{l}(r)$ is derived
from the all-electron valence level with angular momentum $l$ (the reference level
\cite{com:reference-level}) such that

\begin{enumerate}
\item 
The pseudo wavefunction and the all-electron wavefunction correspond to the same 
eigenvalue
      \begin{equation}
      \label{eq:equaleigenvalue}
\epsilon^{\rm ps}_{l} \equiv \epsilon_{nl} \quad,
      \end{equation}
and their logarithmic derivatives (and thus the respective potentials) agree 
beyond a chosen core cutoff radius $r^{\rm c}_{l}$,
\begin{equation}
\label{eq:equallogder}
	\frac{d}{dr} \ln u^{\rm ps}_{l}(\epsilon^{\rm ps}_{l};r)
\rightarrow
	\frac{d}{dr} \ln u_{nl}(\epsilon_{nl};r)
	\quad\mbox{for}\quad r > r^{\rm c}_{l}
\quad.
\end{equation}
\item
The radial pseudo wavefunction has the same amplitude as the all-electron
wavefunction beyond the core cutoff radius,
\begin{equation}
u^{\rm ps}_{l}(\epsilon^{\rm ps}_{l};r) \rightarrow u_{nl}(\epsilon_{nl};r)
\quad\mbox{for}\quad r > r^{\rm c}_{l}
\quad,
\end{equation}
and is normalized, $\int_{0}^{\infty} |u^{\rm ps}_{l}(\epsilon^{\rm ps}_{l};r)|^2 dr
= \int_{0}^{\infty} |u_{nl}(\epsilon_{nl};r)|^2 dr = 1$. This implies the norm-conservation
constraint
\begin{equation}
\label{eq:equalnorm}
\int_{0}^{r'} |u^{\rm ps}_{ l}(\epsilon^{\rm ps}_{l};r)|^2 dr \equiv
\int_{0}^{r'} |u_{nl}(\epsilon_{nl};r)|^2 dr
\quad\mbox{, for}\quad
r' \geq r^{\rm c}_{l}
\quad.
\end{equation}
\item
The pseudo wavefunction contains no radial nodes. In order to obtain a 
continuous pseudopotential which is regular at the origin, it is twice
differentiable and satisfies $lim_{r\rightarrow 0} u^{\rm ps}_{l}(r) \propto r^{l+1}$.
\end{enumerate}

The pseudopotential components then correspond to an inversion of the 
Schr\"odinger equation (\ref{eq:rseq}) for the respective pseudo 
wavefunctions,
\begin{equation}
\label{eq:invert}
V^{\rm ps, scr}_{l}(r) = \epsilon^{\rm ps}_{l} - \frac{l(l+1)}{2 r^{2}}
+ \frac{1}{2 u^{\rm ps}_{l}(r)}
\frac{d^{2}}{d\,r^{2}} u^{\rm ps}_{l}(r)
\quad.
\end{equation}
and become identical to the all-electron potential beyond $r \geq r^{\rm c}_{l}$.

The ansatz for the spatial function $u^{\rm ps}_{l}(r)$ at the energy 
$\epsilon^{\rm ps}_{l}$ requires three free parameters to meet the 
constraints (i) and (ii).
The Hamann scheme is of this ``minimal'' type, while the Troullier-Martins 
scheme introduces additional constraints, namely that the pseudopotential's 
curvature vanish at the origin, $\frac{d^2}{dr^2} V^{\rm ps, scr}_{l}(r)|_{r=0}=0$, 
and that all first four derivatives of the pseudo and the all-electron wavefunction 
agree at the core cutoff radius. Thereby it achieves softer pseudopotentials
for the $2p$, and the $3,4,5d$ valence states of the first row, and of 
the transition metal elements, respectively. For other elements both schemes 
perform rather alike. We refer to the original references \cite{ham89a} 
and \cite{tro91a} for the details about the construction procedure. Here 
we point out that both schemes technically differ somewhat regarding the radii 
where the pseudo wavefunctions and pseudopotentials match their all-electron 
counterparts. The Hamann scheme accomplishes the matching exponentially 
beyond the core cutoff radius 
$r^{\rm c-H}_{l}$, while in the Troullier-Martins scheme this matching 
is exact at and beyond $r^{\rm c-TM}_{l}$. Although in the Hamann scheme 
the core cutoff radii are nominally smaller, $r^{\rm c-H}_{l}/r^{\rm c-TM}_{l} 
\approx 0.25 \ldots 0.75 $, the pseudo wavefunctions and pseudopotentials 
converge to the all-electron wavefunctions and potential within a similar 
range as in the Troullier-Martins scheme.
 
Default values for the core cutoff radii are provided by the program {\sf psgen}.
In the Hamann case it determines the position $r^{\rm Max}_{l}$ of the outermost 
maximum of the all-electron wavefunction and sets $r^{\rm c-H}_{l}=0.6\, r^{\rm Max}_{l}$
if there are core states present with the same angular momentum, and to
$r^{\rm c-H}_{l}=0.4\, r^{\rm Max}_{l}$ otherwise~\cite{ham89a}. In the 
Troullier-Martins case the core cutoff radii are set to values which we have 
tabulated for all elements up to Ag (except the lanthanides). Except for some 
finetuning that might be necessary when the pseudopotentials are used in the 
Kleinman-Bylander form (see Sec.~\ref{sec:kb}) these default values should yield a 
reasonable compromise between pseudopotential transferability and efficiency.

In general, increasing $r^{\rm c}_{l}$ yields softer pseudopotentials, 
which converge more rapidly e.g. in a plane wave basis. These will become
less transferable however as the pseudo wavefunctions become less accurate 
at radii relevant to bonding. When decreasing $r^{\rm c}_{l}$ one has to
keep in mind that $r^{\rm c}_{l}$ has to be larger than the radius 
of the outermost radial node in the respective all-electron orbital. Taking 
$r^{\rm c}_{l}$ too close to such a node results in poor pseudopotentials,
as it is essentially at variance with requiring both a nodeless and at the same 
time norm-conserving pseudo wavefunction. A rugged or multiple-well structure
seen in a plot of the screened pseudopotential and poor scattering properties 
(see Sec.~\ref{sec:transferability}) can indicate such a breakdown.

\subsubsection*{Treating components without bound reference states}

So far we have considered pseudopotentials for bound reference states. On the 
other hand it is quite common in LDA and GGA that unoccupied states in the 
neutral atom are not or only weakly bound, for instance the $3d$-wavefunctions 
for Al, Si, etc.\,. To derive a $l$-component of the pseudopotential in such cases
we follow the generalized norm-conserving procedure proposed by Hamann~\cite{ham89a}: 
At a chosen energy $\epsilon_{l}$ the Schr\"odinger 
equations for the all-electron potential and the pseudopotential are integrated 
from the origin up to a matching radius $r^{\rm match}_{l}$.  
There the logarithmic derivatives and the amplitudes of the pseudo and 
all-electron partial waves are matched, and the ``normalization''
\mbox{
$\int_{0}^{r^{\rm match}_{l}} |u^{\rm ps}_{l}(\epsilon_{l};r)|^2 dr
= \int_{0}^{r^{\rm match}_{l}} |u_{nl}(\epsilon_{l};r)|^2 dr \, := \, 1$}
is enforced. This procedure is analogous to (\ref{eq:equaleigenvalue}) -- 
(\ref{eq:equalnorm}) for bound states. The corresponding pseudopotential 
component results from (\ref{eq:invert}), and merges into the all-electron 
potential for $r > r^{\rm match}_{l}$.
The value of $\epsilon_{l}$ should be chosen in the energy range where the 
valence states are expected to form bands or molecular orbitals, as a 
default, the program {\sf psgen} employs the highest eigenvalue of 
the occupied states, e.g., for Al to $\epsilon_{l=2} = \epsilon_{3p}$. 
Hamann's {\em procedure} for unbound states is not tied to a particular type
of pseudopotential, we use it in constructing Troullier-Martins as well as Hamann
pseudopotential {\em functions}. 
Figure~\ref{fig:al:pot} displays the Hamann-type pseudo wavefunctions and 
pseudopotential components for aluminum. Alternatively, one can derive the 
$l$-components without bound states in the neutral atom from a (separate) 
calculation of an ionized or excited atom which supports bound states for 
these $l$'s~\cite{bac82a}. The above prescriptions should both yield rather 
similar pseudopotentials, consistent with the supposition that pseudopotentials 
are transferable among neutral and excited atomic states.

\subsection{Ionic pseudopotentials, unscreening, and nonlinear core-valence
exchange-correlation}
\label{sec:unscreen}

\begin{figure}[t]
\caption{Graphical output from {\sf psgen} for aluminum.
The left panel shows the valence wavefunctions for the pseudo atom
(solid lines) and the all-electron atom (dashed).
The right panel shows the respective Hamann-type ionic pseudopotential,
using Hamann's procedure to treat the $d$-component
(see Sec.~\ref{sec:screened}) and the LDA.
The legends give the core cutoff radii $r^{\rm c}_l$ (in bohr).
}
\label{fig:al:pot}
\vskip 1em
      \psfrag{x=radius}[c][]{\hspace{-1em}\labelssize $r\quad$ (bohr)}
      \psfrag{y=wavefunctions}[][l]{\labelssize \hspace{8em}$u_l(r)\quad$ (arbitrary scale)}
      \psfrag{ss}{\legendsize$3s$}
      \psfrag{pp}{\legendsize$3p$}
      \psfrag{dd}{\legendsize$3d$}
      \psfrag{rmatch}{\legendsize$r^{\rm match}_{2}$}
      \psfrag{legend1}{\legendsize\makebox[2em][l]{$1s$}$r^{\rm c}_{0} = 1.212$}
      \psfrag{legend2}{\legendsize\makebox[2em][l]{$3s$}}
      \psfrag{legend3}{\legendsize\makebox[2em][l]{$2p$}$r^{\rm c}_{1} = 1.547$}
      \psfrag{legend4}{\legendsize\makebox[2em][l]{$3p$}}
      \psfrag{legend5}{\legendsize\makebox[2em][l]{$3d$}$r^{\rm c}_{2} = 1.547$}
      \psfrag{legend6}{\legendsize\makebox[2em][l]{$3d$}}
      \begin{minipage}{.42\linewidth}
            \epsfig{figure=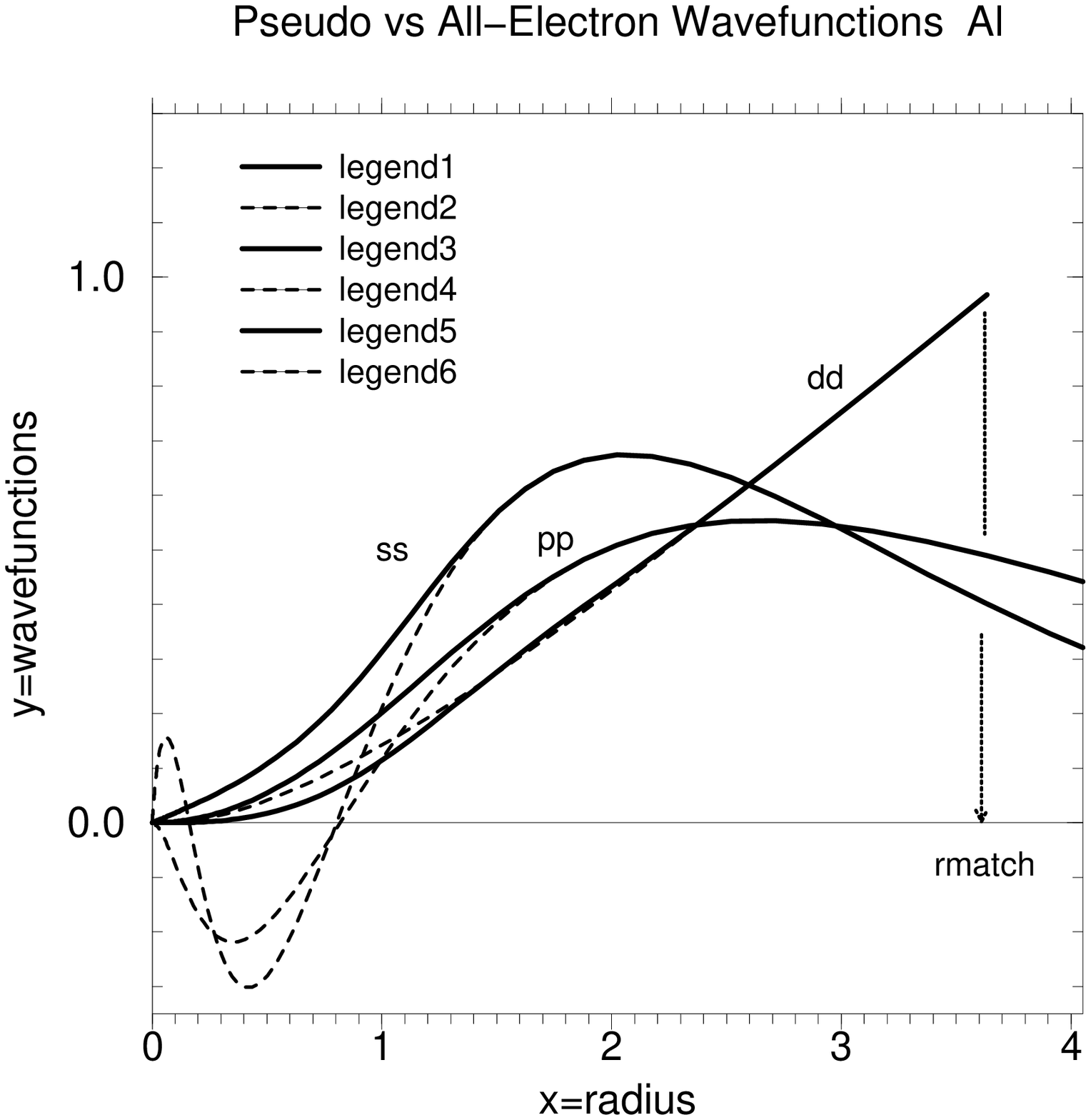,width=\linewidth}
      \end{minipage}
      \hskip 0.1\linewidth
      \psfrag{x=radius}[c][]{\hspace{-1em}\labelssize $r\quad$ (bohr)}
      \psfrag{y=potentials}[bl]{\labelssize $V^{\rm ps}_{l}(r)\quad$ (hartree)}
      \psfrag{ss}{\legendsize$3s$}
      \psfrag{pp}{\legendsize$3p$}
      \psfrag{dd}{\legendsize$3d$}
      \psfrag{Coulomb}{\legendsize$\displaystyle{-\frac{3}{r}}$}
      \psfrag{legend1}{\legendsize\makebox[1.5em][l]{0}$r^{\rm c}_{0} = 1.212$}
      \psfrag{legend2}{\legendsize\makebox[1.5em][l]{1}$r^{\rm c}_{1} = 1.547$}
      \psfrag{legend3}{\legendsize\makebox[1.5em][l]{2}$r^{\rm c}_{2} = 1.547$}
      \begin{minipage}{.42\linewidth}
            \centering
            \epsfig{figure=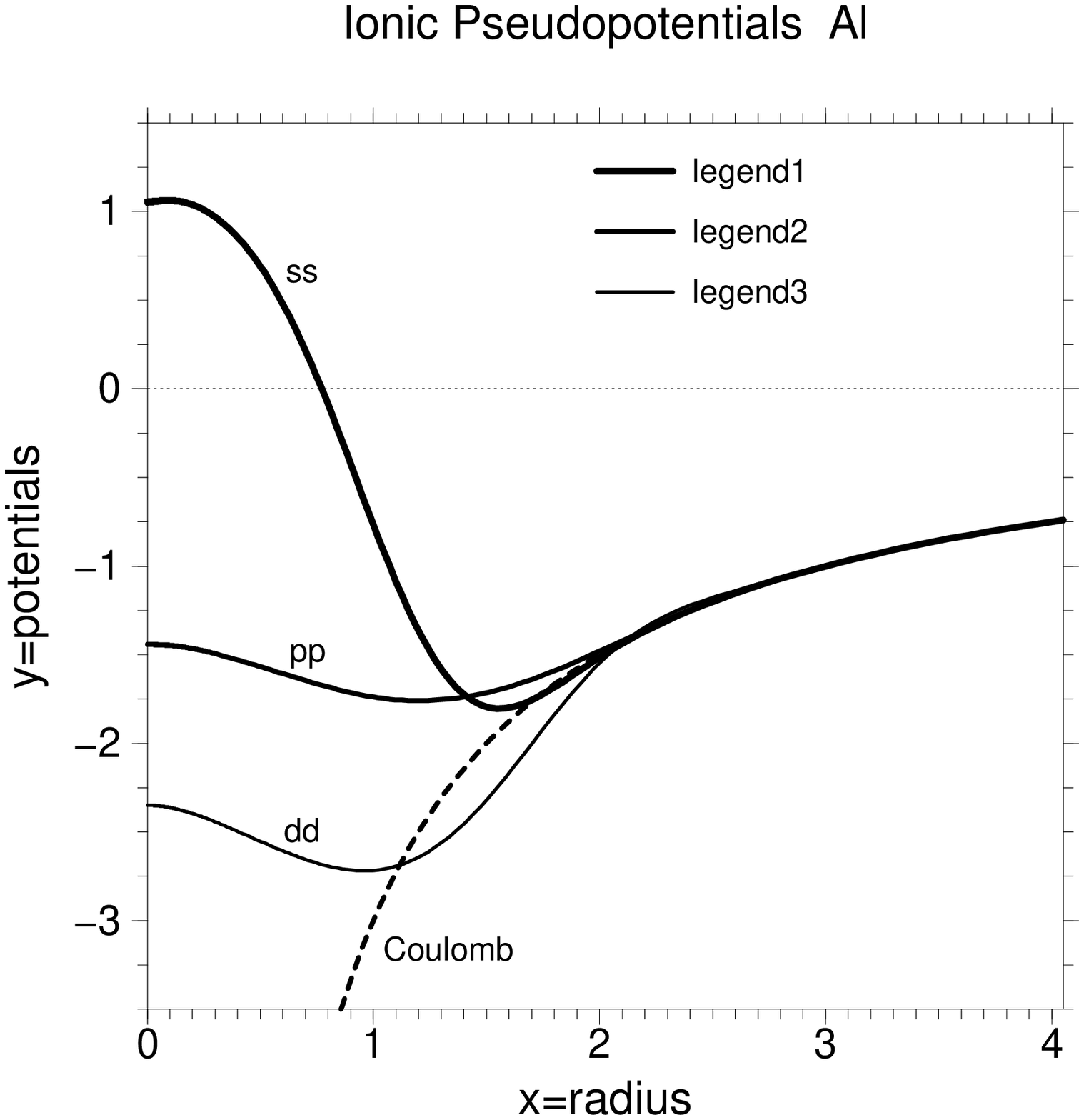,width=\linewidth}
      \end{minipage}
\vskip 1em
\end{figure}

The final ionic pseudopotentials are determined by subtracting from 
the screened pseudopotentials  
the electrostatic and the exchange-correlation screening contributions 
due to the valence electrons, 
\begin{equation}
\label{eq:ionic:pp}
V^{\rm ps}_{l}(r) = V^{\rm ps,scr}_{l}(r) - V^{\rm H}[\dens^{\rm ps}_0;r] 
						  - V^{\rm XC}[\dens^{\rm ps}_0;r]
\quad\mbox{with}\quad
\dens^{\rm ps}_0(r) = \frac{1}{4\pi}\sum_{l=0} f_{l} 
\left|
\frac{u^{\rm ps}_{l}(\epsilon^{\rm ps}_{l};r)}{r}
\right|^2
\quad.
\end{equation}
The valence electron density is evaluated from the atomic pseudo wavefunctions,
with the same occupancies $f_{l}$ as for the all-electron valence 
states of Sec.~\ref{sec:atomic}. Figure~\ref{fig:al:pot} shows the ionic pseudopotentials 
for aluminum as an example. With the pseudopotentials as defined by Eq.~(\ref{eq:ionic:pp})
the electronic total energy of an arbitrary system reads
\begin{equation}
\label{eq:etot:ps}
E^{\rm tot} = \sum_i f_i \langle \psi_i | \hat T + \hat V^{\rm ps} | \psi_i \rangle
+ E^{H}[\dens^{\rm ps}] + E^{\rm XC}[\dens^{\rm ps}]
\quad\mbox{with}\quad
\dens^{\rm ps}(\rvec) = \sum_i f_i |\psi_i(\rvec)|^2
\quad.
\end{equation}

Here, $E^{\rm XC}$ refers to the exchange-correlation interaction between the valence 
electrons themselves. That between the valence and the core electrons is included in 
the pseudopotential energy, as a term which depends linearly on the valence density 
$\dens^{\rm ps}$. Although $E^{\rm XC}$ is a nonlinear functional of the total electron 
density $\dens$,
the above ``linearization'' of its core-valence contribution is a usual and mostly 
adequate approximation for calculations within both LDA and GGA~\cite{fuc98a}. However, an 
explicit account of the core-valence nonlinearity of $E^{\rm XC}$ is sometimes required, 
for instance in studies involving alkali metal atoms or within spin-density functional 
theory. This is accomplished by restoring the dependence of $E^{\rm XC}$ (and $V^{\rm XC}$) 
on the total electron density, i.e. adding the (atomic) core density to the valence 
density in the argument of $E^{\rm XC}$. In turn the ionic pseudopotentials (\ref{eq:ionic:pp}) 
are redefined. Rather than the full core density, it suffices to add a so 
called partial core density $\tilde\dens^{\rm core}_{0}$, as suggested by 
Louie {\em et al.}~\cite{lou82a}. It reproduces the full core density 
$\dens^{\rm core}_0$ (\ref{eq:core-density}) outside a chosen cutoff radius 
$r^{\rm nlc}$ but is a smoother function inside, which enables its later use
together with plane waves. We use a polynomial
\begin{equation}
\tilde \dens^{\rm core}_0(r)
= \left\{
\begin{array}{lll}
\dens^{\rm core}_0(r)  & \mbox{for} & \quad r \geq r^{\rm nlc} \,\,,
\\
c_0 + \sum_{i=3}^{6} c_i r^i & \mbox{for} & \quad r < r^{\rm nlc}\,\,, 
						\mbox{with}\quad \tilde \dens^{\rm core}_0(r) <
						\dens^{\rm core}_0(r) \quad, 
\end{array}
\right.
\end{equation}
where we take the coefficients $c_i$ such that $\tilde \dens^{\rm core}_0(r)$ 
has zero slope and curvature at the origin, decays monotonically, and joins 
the full core density continuously up to the third derivative.
In this way smooth exchange-correlation potentials result for the LDA 
as well as for the GGAs. The resulting nonlinear core-valence exchange-correlation 
scheme uses the redefined ionic pseudopotentials 
\begin{equation}
V^{\rm ps}_{l}(r) \rightarrow  V^{\rm ps,scr}_{l}(r) - V^{\rm H}[\dens^{\rm ps}_0;r]
                   - V^{\rm XC}[\dens^{\rm ps}_0+\tilde\dens^{\rm core}_0;r]
\quad
\end{equation}
instead of Eq.~(\ref{eq:ionic:pp}). The according total energy expression is
\begin{equation}
\label{eq:nlcv-xc-etot}
E^{\rm tot} \rightarrow \sum_i f_i \langle \psi_i | \hat T + \hat V^{\rm ps} | \psi_i \rangle
+ E^{H}[\dens^{\rm ps}] + E^{\rm XC}[\dens^{\rm ps}+\tilde\dens^{\rm core}_0]
\quad,
\end{equation} 
where, in a poly-atomic system, $\tilde\dens^{\rm core}_0$ stands for 
the superimposed partial core densities of the constituent atoms.

Whether nonlinear exchange-correlation plays a significant role can be 
readily decided by a comparison of two calculations with and without a 
partial core density. As a rule, it becomes more important to the left in the periodic table,
i.e. with fewer electrons in the valence shell (alkali metals), and with increasing
atomic number, i.e. the farther the upper core orbitals extend into
the tail of the valence density (e.g. Zn, Cd). Of course if the uppermost semicore
states hybridize with ``true'' valence states they have to be treated as valence 
states in the first place, like the $3,4,5d$-states in all transition and noble metals, 
but also the $3d$-states in Ca. This also applies for the $3d$- and $4d$-states 
of Ga and In in GaN and InN, where these interact markedly with the N atoms' 
$2s$-states in GaN and InN, while treating them in the core is indeed adequate 
in, say, GaAs or InAs. 

There remains the question what value to
pick for the core cutoff radius $r^{\rm nlc}$. It is reasonable to set 
$r^{\rm nlc}$ to about the radius where the full core density drops below
the valence electron density. The utility {\sf psgen} determines this 
``equi-density radius'' and also outputs the core and valence densities. 
Choosing $r^{\rm nlc}$ smaller makes it more demanding to represent the 
partial core, say, in a plane wave basis, without further enhancing the 
pseudopotential's performance.  On the other hand, too large a value 
for $r^{\rm nlc}$ might be insufficient to capture the nonlinearities in 
the desired way. Of course, a systematic convergence test can be carried 
out by inspecting the variation with respect to $r^{\rm nlc}$ of some simple 
system's properties, e.g., the crystal equilibrium volume.

The GGA, unlike the LDA, occasionally gives rise to short-ranged oscillations
in the ionic pseudopotentials, particularly when Eq.~(\ref{eq:ionic:pp}) is
employed together with the PW91 GGA, but less so when a partial core density
is included. These features may be traced to the behavior of 
$V^{\rm XC-GGA}[\dens^{\rm ps}_0;r]$ subtracted out in the unscreening, and 
are certainly artifacts of the GGA. 
From the viewpoint of plane wave calculations they correspond to an energy
regime where the kinetic energy dominates all other contributions and are,
therefore, physically unimportant. A ``controlled'' smoothing is
then implied already by the plane wave basis cutoff which, in our
experience, can be chosen very similar to that for the LDA. Accordingly
we do not apply any smoothing of the GGA pseudopotentials in real space.
 
\subsection{Transferability and convergence behavior}
\label{sec:transferability}

{\em Transferability considerations} --
By construction, a pseudopotential reproduces the valence states 
of the free atom in the reference configuration. In applications,
it needs to perform correctly in a wide range of different environments, 
say free molecules and bulk crystals, ``predicting'' the same 
valence electronic structure and total energy differences that an 
all-electron method would deliver. This transferability depends 
critically on (a) the choice of the core cutoff radii (see Sec.~\ref{sec:screened}), 
(b) the linearization of core-valence exchange-correlation 
(Sec.~\ref{sec:unscreen}), (c) the frozen-core approximation 
underlying the pseudopotential's construction 
(Sec.~\ref{sec:unscreen}), and (d) the transformation of the 
semilocal into the fully separable form of the pseudopotential 
(discussed in Sec.~\ref{sec:kb}). 

Clearly, the transferability of the pseudopotentials should be 
carefully verified before embarking on extensive computations.
Below we discuss several simple tests to be performed for a
free atom. These tests assume spherical symmetry, and do not
mimick non-spherical environements. It is therefore 
good practice to also cross check -- for some properties of 
simple test systems, e.g., bond lengths in molecules or a bulk crystals 
-- the results from one's pseudopotential calculation with all-electron 
results, which often are available from the literature.
The compared data have to refer to the same approximation for 
exchange-correlation, of course.  Note that a cross check against 
experimental data alone might be less conclusive, since the LDA or 
GGA themselves lead to systematic deviations between theoretical 
and experimental figures which need to be distinguished from errors
rooted in the pseudopotentials themselves.

\begin{figure}[t]
\caption{Transferability tests of the aluminum potential shown in
Figure~\protect{\ref{fig:al:pot}}. Left panel: Logarithmic derivatives
for the semilocal (dashed lines) and fully separable pseudopotentials (dot-dashed)
vs. their all-electron counterparts (solid), as output by {\sf pswatch}.
Right panel: Error of the atomic excitation energies (open symbols)
and the level spacing $\epsilon_{3s}-\epsilon_{3p}$ (filled symbols)
calculated for the pseudo atom and the frozen-core all-electron atom.
}
\label{fig:al:trans}
\vskip 1em
      \psfrag{x=energy}[c][][\skala]{\labelssize \hspace{2em}energy $\quad$ (hartree)}
      \psfrag{y=logderivative}[c][][\skala]{\labelssize logarithmic derivative \,\,(arbitrary scale)}
      \psfrag{dd}[][][\skala]{\legendsize $d$}
      \psfrag{pp}[][][\skala]{\legendsize $p$}
      \psfrag{ss}[][][\skala]{\legendsize $s$}
      \psfrag{subtitle}[c][][\skala]{\titlessize\hspace{1.5em}Al $\quad r^{\rm diag} = 2.99$~bohr $\quad l_{\rm loc} = 2$}
      \begin{minipage}{.42\linewidth}
      	\epsfig{figure=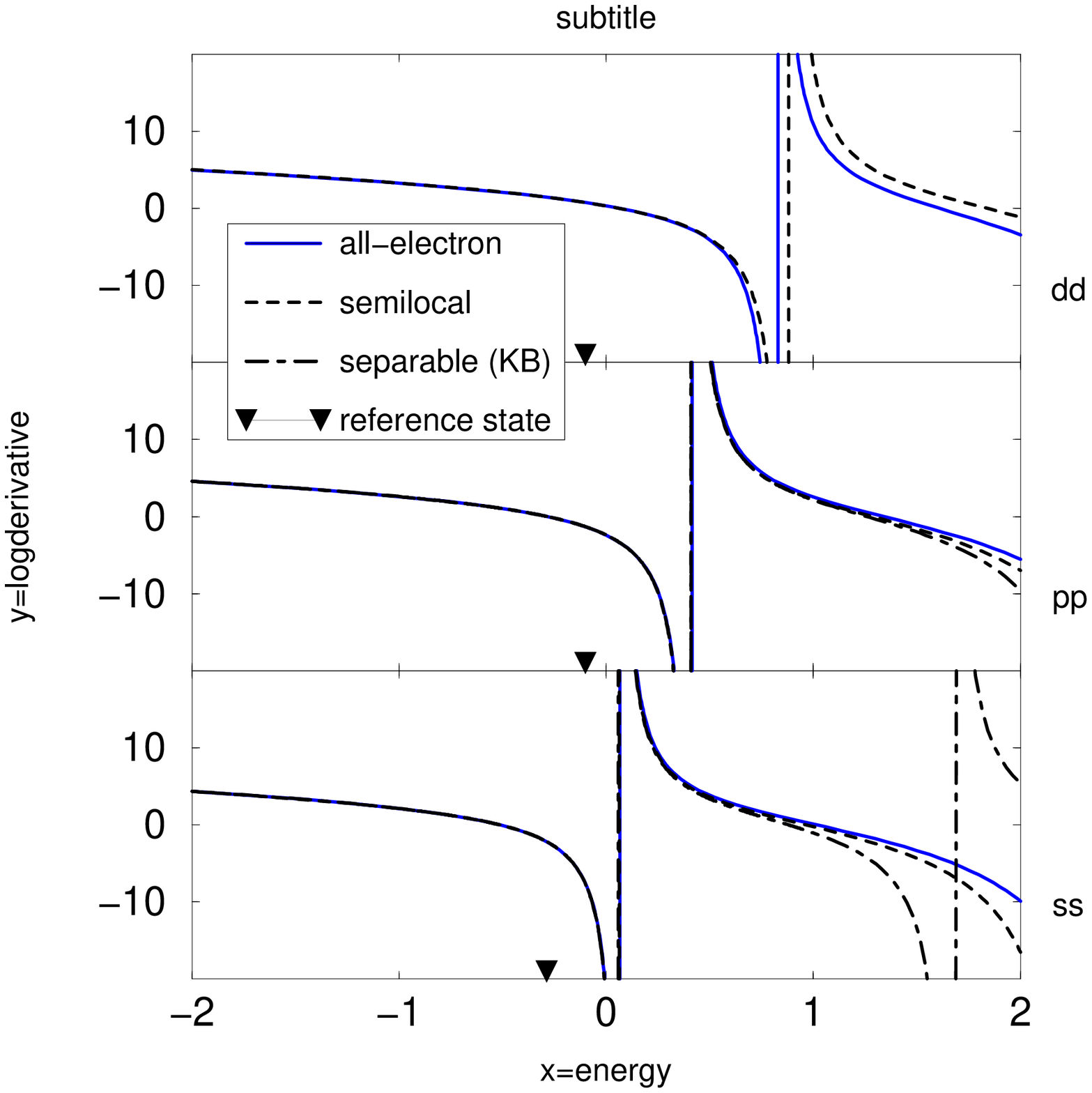,width=\linewidth}
      \end{minipage}
	\hskip 0.1\linewidth
      \psfrag{x=configuration}[c][][\skala]{\labelssize \hspace{1em} occupancy}
      \psfrag{y=error}[c][][\skala]{\labelssize error (meV)}
      \psfrag{3s}[][][\skala]{\legendsize$3s$:}
      \psfrag{3p}[][][\skala]{\legendsize$3p$:}
	\begin{minipage}{.42\linewidth}
      	\epsfig{figure=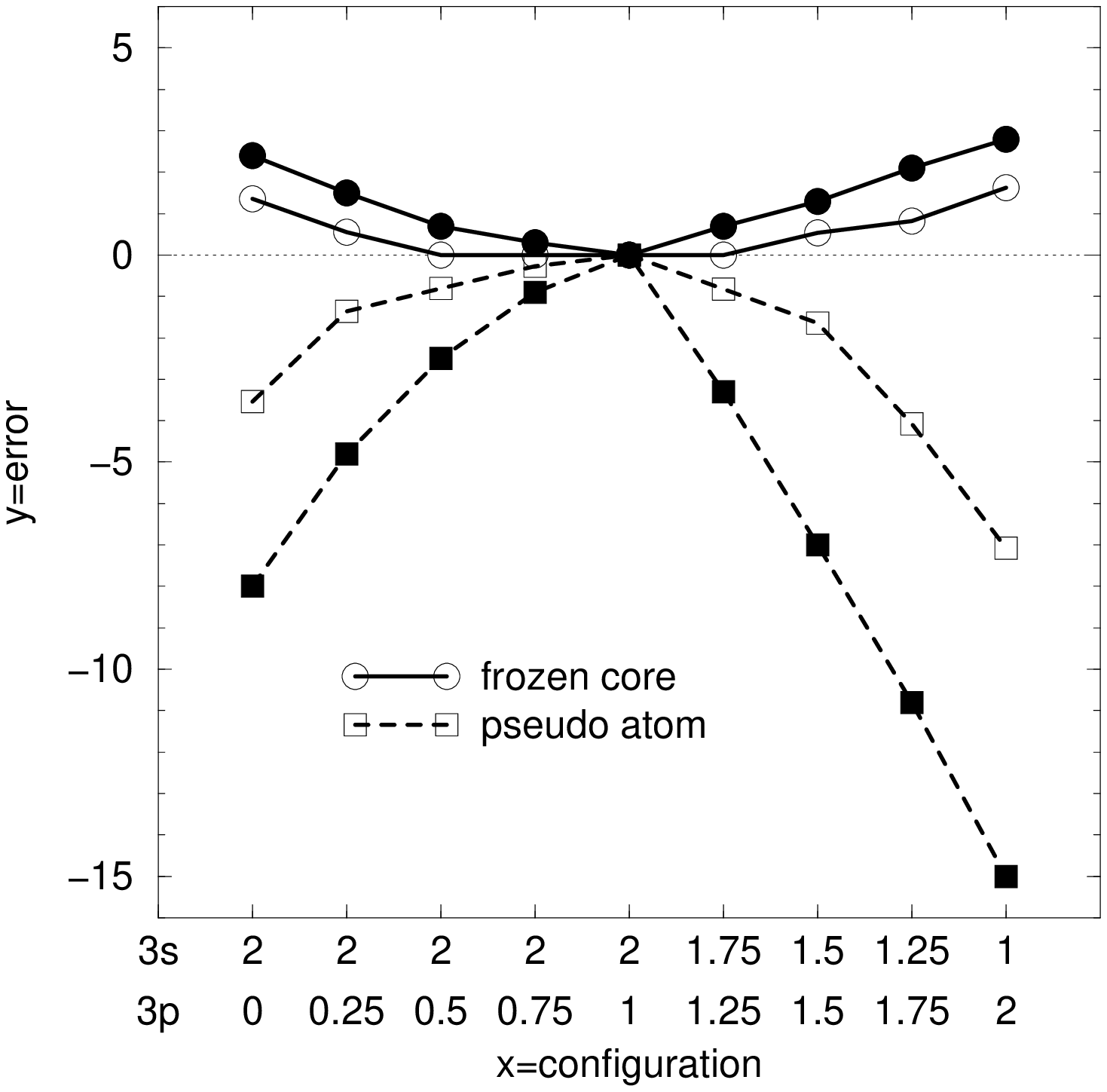,width=\linewidth}
	\end{minipage}
\vskip 1em
\end{figure}

{\em Test of scattering properties} --
As a first test, we establish that the logarithmic derivatives of 
the radial wavefunctions,
\begin{equation} 
D_{l}(\epsilon,r^{\rm diag}) = \left. \frac{d}{dr} 
\ln u_{l}(\epsilon;r)\right|_{r=r^{\rm diag}}\quad, 
\end{equation}
agree for the pseudo and the
all-electron atom, as a function of the energy at some 
diagnostic radius $r^{\rm diag}$ outside the core region, say half of a 
typical interatomic distance. 
The norm-conservation feature ensures that the logarithmic 
derivatives for the pseudopotential (semilocal or fully separable) 
are correct to first order around the reference energies $\epsilon^{\rm ps}_l$,
by Eqs.~(\ref{eq:norm1}) and ~(\ref{eq:equalnorm}),
\begin{equation}
\left.
\frac{d}{d\epsilon}\frac{d}{dr} \ln u_{nl}(\epsilon; r)
\right|_{\epsilon = \epsilon^{\rm ps}_{l}}
=
\left.
\frac{d}{d\epsilon}\frac{d}{dr} \ln u^{\rm ps}_{l}(\epsilon; r)
\right|_{\epsilon = \epsilon^{\rm ps}_{l}}
\,\,\mbox{, with}\quad r \geq r^{\rm c}_{l}
\quad.
\end{equation}
More generally the pseudopotential and all-electron logarithmic 
derivatives should closely agree over the whole range of energies where 
the valence states are expected to form Bloch bands or molecular 
orbitals, say $\pm 1$ hartree around the atomic valence 
eigenvalues. 
Figure~\ref{fig:al:trans} shows, as a typical case, the logarithmic 
derivatives for an aluminum pseudopotential, evaluated with this 
package's utility {\sf pswatch}.
Note that the logarithmic derivatives are evaluated for the screened
potentials of the atomic reference configuration, and will look 
different for an excited configuration.

{\em Test of excitation energies} --
Next we verify that the pseudopotential reproduces the all-electron 
results for the atomic excitation or ionization energies, given by
\begin{equation}
E^{\rm M}_{ba} = E^{\rm tot-M}(\lbrace f_i^{b}\rbrace) - 
				E^{\rm tot-M}(\lbrace f_{i}^{a}\rbrace)
\quad \mbox{with M:}\left\{
\begin{array}{l}
\mbox{pseudopotential calculation (PS),}\\
\mbox{all-electron calculation (AE),} \\
\mbox{all-electron frozen-core calc. (FC),}
\end{array}\right.
\end{equation}
where $\lbrace f_i^{b,a} \rbrace$ denote the orbital occupancies in
the excited and in the ground state configuration, respectively.
The various expressions for $E^{\rm tot-M}$ are collected in the Appendix.
This test seeks to emulate the cost in total energy associated with the 
orbital hybridization in different environments (``promotion energies''). 
The errors due to the use of pseudopotentials,
$\Delta E^{\rm PS}_{ba}=E^{\rm PS}_{ba} - E^{\rm AE}_{ba}$ should be compared to
$\Delta E^{\rm FC}_{ba}=E^{\rm FC}_{ba} - E^{\rm AE}_{ba}$, the errors 
that result from an all-electron calculation within the frozen-core approximation.
There one recalculates just the valence states (with the full nodal structure), but 
keeps the core density fixed as in the ground state configuration 
$\{f_i^{a}\}$. Due to 
the variational principle, frozen-core calculations yield 
higher total energy values than unrestricted all-electron 
calculations where the core is allowed to relax, 
$E^{\rm tot-FC}(\{f^{b}_i\}) \geq E^{\rm tot-AE}(\{f^{b}_i\})$,
the equality holding only for $\{f_i^{a}\}$.
In this test a transferable pseudopotential should perform 
with the same accuracy as the frozen-core all-electron calculation,
$|\Delta E^{\rm PS}_{ba}| \simeq |\Delta E^{\rm FC}_{ba}|$,
where the errors are typically of opposite sign for linearized 
core-valence exchange-correlation, and essentially the same 
if nonlinear core-valence exchange-correlation is taken into 
account. Typically, say for the first ionization potential,
the errors do not exceed a few ten meV. Figure~\ref{fig:al:trans} 
shows the excitation energy error for an aluminum pseudopotential.

{\em Test of level changes (chemical hardness)} -- 
As a last point we verify that the orbital eigenvalues for 
the pseudo atom closely track those of the all-electron atom 
when the orbital configuration changes. This test provides a
discrete sampling of the atomic hardness matrix,
$H_{ij}^{b} := \frac{\partial}{\partial f_i} \epsilon_j(\{ f_k^{b} \}) =$
$\frac{\partial^2}{\partial f_i \partial f_j} E^{\rm tot}(\{ f_k^{b} \})$
which ultimately governs the changes of the one-particle levels ~\cite{tet93a,fil95a}
and, by Janak's theorem~\cite{jan78}, of the total energy as functions of 
the orbital occupancies. Like for 
the excitation energies, the eigenvalues from the pseudopotential 
calculation are expected to agree with the all-electron results to 
within the error found for the frozen-core approximation, as
seen in Fig.~\ref{fig:al:trans}.

\subsubsection*{Convergence behavior}

\begin{figure}[t]
\caption{
Convergence of the total energy in pseudopotential plane wave
calculations of bulk GaAs, diamond, and fcc-Cu. Solid lines show the
absolute total energy error (per electron) expected from the atomic pseudo wavefunctions
using the estimate Eq.~(\protect{\ref{eq:ekin-criterion}}). Open circles
show the actual values with respect to completely converged calculations.
The weights $\{w_{l=0},w_{l=1},w_{l=2}\}$ taken in Eq. (\protect{\ref{eq:ekin-criterion}})
were $\{2,1,0\}$ (Ga $4s^2 4p^1 4d^0$), $\{2,3,0\}$ (As $4s^2 4p^3 4d^0$),
$\{2,2,0\}$ (C $2s^2 2p^2 3d^0$), and $\{1,0,10\}$ (Cu $3d^{10} 4s^1 4p^0$).
Troullier-Martins pseudopotentials were employed with core cutoff
radii $\{r^{c}_{l=0},r^{c}_{l=1},r^{c}_{l=2}\}$ (in bohr) of $\{2.2,2.3,2.5\}$ (Ga),
$\{2.1,2.2,2.4\}$ (As), $\{1.5,1.5,1.5\}$ (C), and $\{2.1,2.3,2.1\}$ (Cu).
}
\label{fig:ekin-criterion}
\vskip 1em
     \psfrag{x=cutoff}[c][][1.0]{\labelssize plane wave cutoff energy \quad (Ry)}
     \psfrag{y=error}[c][][1.0]{\labelssize total energy error per electron\,\, (eV)}
     \centering
     \epsfig{figure=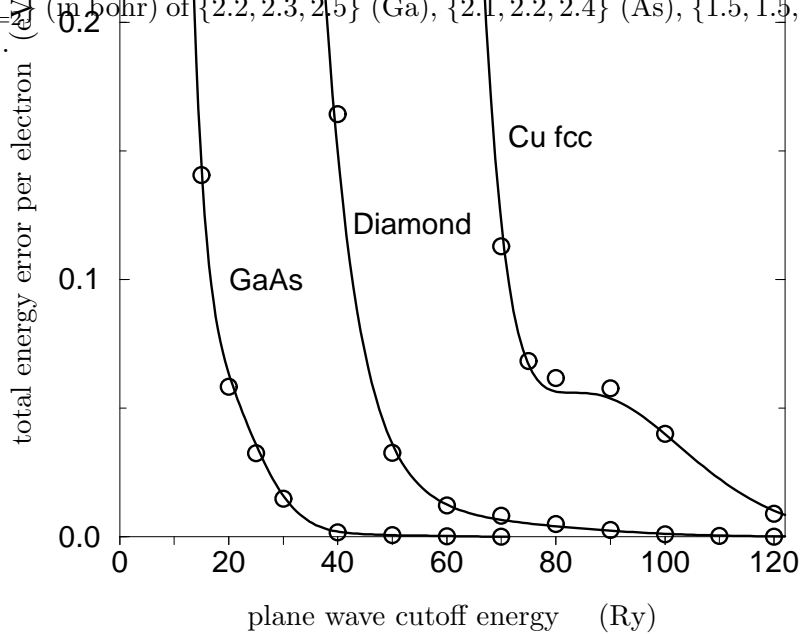,width=0.42\linewidth}
\vskip 1em
\end{figure}
Checking the convergence behavior of the pseudopotentials with 
respect to the basis size is a basic task in calculating realistic
systems, where the goal is to achieve convergence in total energy 
differences, e.g. binding energies, rather than the total energy 
itself. To provide some guidance in picking the smoothest 
out of a set of some similarly transferable pseudopotentials,
and in choosing an initial estimate for a plane wave basis, the 
program {\sf pswatch} evaluates the momentum space convergence of 
the atomic pseudo wavefunctions' kinetic energy. This provides a measure 
for the (absolute) total energy convergence in applications~\cite{kre94a}. 
Using the Fourier decomposition of the pseudo wavefunctions, 
\begin{equation}
u^{\rm ps}_{l}(k)
= \sqrt{\frac{2}{\pi}}\int_0^{\infty} kr j_{l}(kr) u^{\rm ps}_{l}(r) \,dr
\quad,
\end{equation}
consider the error of the momentum space value of the kinetic energy as 
a function of the plane wave cutoff energy $E^{\rm pw}$ corresponding
to the cutoff momentum $k^{\rm pw} = \sqrt{2E^{\rm pw}}$,
\begin{equation}
\label{eq:ekin-reciprocal}
\Delta E^{\rm kin}_{l}(k^{\rm pw})
= \int_{0}^{k^{\rm pw}} 
      \frac{k^2}{2}\, |u^{\rm ps}_{l}(k)|^2 
      dk
-
\int_0^{\infty} u^{\rm ps}_{l}(r)
      \left[
            - \frac{1}{2}\frac{d^2}{dr^2} + \frac{l(l+1)}{2r^2}
      \right] 
      u^{\rm ps}_{l}(r) dr
\quad.
\end{equation}
The program {\sf pswatch} computes the cutoff energies to achieve 
$\Delta E^{\rm kin}_{l} \leq 100, 10, 1\, \mbox{meV}$. 
For real systems a converged calculation typically corresponds 
to a plane wave cutoff that meets the criterion 
$\Delta E^{\rm kin}_{l}(k^{\rm pw}) \leq 100\,\,\mbox{meV}$
for the pseudo atom~\cite{com:ekin-criterion}. A corresponding
estimate of the (absolute) convergence of the total energy in 
poly-atomic systems is then given by the average of the errors 
$\Delta E^{\rm kin}_{l}$ over all states and atoms,
\begin{equation}
\label{eq:ekin-criterion}
E^{\rm tot}(k^{\rm pw}) - E^{\rm tot}(\infty) 
\simeq
\sum_{l} w_{l} \Delta E^{\rm kin}_{l}(k^{\rm pw})
\Bigl|_{{\rm atom}\,1} + \ldots \Bigl|_{{\rm atom}\,2} + \ldots
\quad,
\end{equation}
with the orbital weights $w_l$. Figure~\ref{fig:ekin-criterion} illustrates 
this estimate.

\subsection{Usage of pseudopotentials: Separation of long-ranged local and 
short-ranged nonlocal components}
\label{sec:use-local}

Operating with the ionic pseudopotential on the wavefunctions in a 
poly-atomic system requires their projection onto an angular momentum 
expansion, the second term of Eq. (\ref{eq:semilocal-pseudopotential}), 
which causes the predominant computational effort in calculating larger systems.
As all components $V^{\rm ps}_{l}(r)$ at large $r$ reduce to the 
ionic Coulomb potential, $-Z^{\rm ion}/r$, getting independent of $l$.
It is thus expedient to express the pseudopotential as a 
multiplicative local potential plus only a few $l$-dependent, short-ranged
``corrections'' for $l < l_{\rm max}$,
\begin{eqnarray}
\!\!\!\langle\rvec | \hat V^{\rm ps} | \rvec'\rangle
&=&
\langle\rvec |\hat V^{\rm loc} + \delta\hat V^{\rm sl}|\rvec'\rangle
\\
&\rightarrow&
V^{\rm ps}_{l_{\rm loc}}(r)\delta(\rvec-\rvec')
+\sum_{l=0}^{l_{\rm max}}\!\sum_{m=-l}^{l}\!
Y^{\ast}_{lm}(\Omega_{\rvec}) \delta V^{\rm sl}_{l}(r) \frac{\delta(r-r')}{r^2}
Y_{lm}(\Omega_{\rvec'}),
\end{eqnarray}
where the local potential is taken as one of the semilocal pseudopotential
components, $V^{\rm loc}(r) = V^{\rm ps}_{l=l_{\rm loc}}(r)$, and 
$\delta V^{\rm sl}_{l}(r) = V^{\rm ps}_{l}(r) - V^{\rm ps}_{l_{\rm loc}}(r)$
vanishes beyond $r^{\rm c}_{l}$.
Typically one truncates the expansion at $l_{\rm max}=2$, whereby the 
core effects on the valence states with the same symmetries as the core
states are accounted for. The number of projections is reduced most 
if one picks the $l_{\rm loc} = l_{\rm max}$ component as local potential 
$V^{\rm loc}(r) = V^{\rm ps}_{l_{\rm loc}}(r) = V^{\rm ps}_{l_{\rm lmax}}(r)$.
Most preferably $l_{\rm loc}=l_{\rm max}=2$, the common practice for $sp$-type 
materials like silicon.  We note that the local component of the pseudopotential 
in principle has to reproduce the scattering properties in the higher angular 
momentum channels with $l>l_{\rm max}$. Further aspects on the choice of the
local potential come into play when one uses the fully separable form 
of the pseudopotential as discussed in the next Section.

\subsection{Fully separable pseudopotentials and ghost states}
\label{sec:kb}

Actual electronic structure codes mostly use the ionic pseudopotential
in the fully separable form of Kleinman and Bylander (KB)~\cite{kle82a}
\begin{eqnarray}
\langle \rvec | \hat V^{\rm ps} |\rvec'\rangle
\quad &=& \quad 
\langle \rvec | \hat V^{\rm loc} + \delta\hat V^{\rm KB} |\rvec'\rangle
\label{eq:kb-potential-op}
\\
&\rightarrow& \quad
V^{\rm ps}_{l_{\rm loc}}(r)\delta(\rvec-\rvec')
+ \sum_{l = 0}^{l_{\rm max}}
\sum_{m=-l}^{l}
\langle \rvec |\chi^{\rm KB}_{lm} \rangle E^{\rm KB}_{l} \langle \chi^{\rm KB}_{lm}| \rvec'\rangle
\quad, 
\label{eq:kb-potential-r}
\end{eqnarray}
where the short-ranged second term is a fully nonlocal rather than a 
semilocal operator in $\rvec$-space. Starting from a semilocal pseudopotential,
the corresponding KB-pseudopotential is constructed such that it yields
exactly the same pseudo wavefunctions and energies. This is accomplished
with the projector functions
\begin{equation}
\label{eq:ukb}
\langle \rvec | \chi^{\rm KB}_{lm} \rangle
= 
\frac{1}{r}
\chi_{l}(r) Y_{lm}(\Omega_{\rvec}) 
=
\frac{1}{r}
\frac{
u^{\rm ps}_{l}(r) \delta V^{\rm sl}_{l}(r)
}{
|| u^{\rm ps}_{l} \delta V^{\rm sl}_{l}||^{1/2}
}
\,Y_{lm}(\Omega_{\rvec})
\quad,
\end{equation}
and the KB-energies 
\begin{equation}
\label{eq:ekb}
E^{\rm KB}_{l} = 
\frac{
|| u^{\rm ps}_{l} \delta V^{\rm sl}_{l}||
}{
\langle \psi^{\rm ps}_{lm} | \chi^{\rm KB}_{lm} \rangle 
}
\quad,
\end{equation}
with $||u^{\rm ps}_{l}\delta V^{\rm sl}_{l}|| = 
\int_{0}^{\infty} u^{\rm ps}_{l}(r) \delta V^{\rm sl}_{l}(r)^2 u^{\rm ps}_{l}(r) dr$
~\cite{com:projectors}.
The KB-energy measures the strength of a nonlocal component relative 
to the local part of the pseudopotential, where the sign is determined by the KB-cosine 
$-1 < \langle \psi^{\rm ps}_{lm}|\chi^{\rm KB}_{lm} \rangle < 1$.
Note that $\chi_{l}(r) \rightarrow 0$ as $\delta V^{\rm sl}_l(r) \rightarrow 0$ 
outside the core region.
In poly-atomic systems the KB-form drastically reduces the effort 
of operating with the pseudopotentials on the wavefunctions, 
which causes the dominant computational workload in particular for 
larger systems: for an $N$-dimensional basis 
$\{ |G_{i}\rangle\}$, the semilocal form requires the evaluation and 
storage of $\sim (N^2+N)/2$ matrix elements 
$\langle G_{i}| \delta\hat V^{\rm sl}_{l}| G_{j} \rangle$, 
with the KB-form these factorize, requiring just $\sim N$ projections 
$\langle G_{i}| \chi^{\rm KB}_{lm}\rangle$ and simple multiplications 
$\langle G_{i}| \chi^{\rm KB}_{lm}\rangle E^{\rm KB}_{l} \langle \chi^{\rm KB}_{lm}|G_{j}\rangle$.

In transforming a semilocal to the corresponding KB-pseudopotential
one needs to make sure that the KB-form does not lead to unphysical 
``ghost'' states at energies below or near those of the physical 
valence states as these would undermine its transferability.
Such spurious states can occur for specific (unfavorable) choices 
of the underlying semilocal and local pseudopotentials. They are 
an artifact of the KB-form's nonlocality by which the nodeless 
(reference) pseudo wavefunctions need not be the lowest possible 
eigenstate, unlike for the semilocal form~\cite{gon91a}.
Formally the KB-form can be generalized to a series expansion of 
a fully nonlocal pseudopotential which avoids ghost states by 
projecting onto additional reference states in Eq.~(\ref{eq:kb-potential-r})
~\cite{blo89a}, yet at the cost of somewhat increasing the computational
labor. In practice transferable ghost-free KB-pseudopotentials 
are readily obtained by a proper choice of the local component ($l_{\rm loc}$)
and the core cutoff radii ($r^{\rm c}_l$) in the basic semilocal pseudopotentials. 
In the next paragraphs we discuss how to identify and remove ghost states 
using the utility {\sf pswatch}.

First of all a ghost state is indicated by a marked deviation of the
logarithmic derivatives of the KB-pseudopotential from those of the 
respective semilocal pseudopotential or all-electron potential.
Note that such features can be delicate to detect if the energy
sampling is too coarse to resolve them or if they occur below the
inspected energy interval. The program {\sf pswatch} provides the 
logarithmic derivatives on an adaptive energy mesh, integrating the 
nonlocal radial Schr\"odinger equation for the KB-pseudopotential,
\begin{equation}
\label{eq:nl:scheq}
\left[
-\frac{1}{2}\frac{d^2}{dr^2} + \frac{l(l+1)}{2r^2} + V^{\rm ps,scr}_{l_{\rm loc}}(r)
- \epsilon
\right]
u^{\rm ps}_{l}(\epsilon; r)
+ \chi_{l}(r)E_{l}
  \int_{0}^{\infty}\!\chi_{l}(r') u^{\rm ps}_{l}(\epsilon; r')\, dr'
= 0
\quad,
\end{equation}
with $E_l \equiv E^{\rm KB}_l$, from the origin up to a diagnostic 
radius, as described in Ref.~\cite{gon91a}. 

Secondly, ghost states are revealed in a comparison of the atomic 
bound state spectra for the semilocal and the KB-pseudopotentials~\cite{khe95b}, 
i.e., as extra levels not present in the true valence spectrum.
The program {\sf pswatch} solves the respective eigenvalue problem, 
Eq.~(\ref{eq:nl:scheq}), using Laguerre polynomials as a numerically
convenient basis~\cite{arf85a}.

Thirdly, ghost states below the valence states are identified by
a rigorous criterion by Gonze {\it et al.}~\cite{gon91a}. The
program {\sf pswatch} evaluates this criterion, here we summarize
its key points.
For a given angular momentum, consider the spectrum 
$\lbrace\tilde\epsilon_{i}(E_{l})\rbrace$ of the nonlocal 
Hamiltonian associated with the KB-form, Eq.~(\ref{eq:nl:scheq}),
as a {\em function} of the strength parameter $E_l$. Comparing 
to the spectrum resulting from the local part of the (screened) 
pseudopotential alone, $\lbrace\tilde\epsilon_{i}(E_{l} \equiv 0)\rbrace$,
the eigenvalues always satisfy the sequences~\cite{gon91a}, 
\begin{eqnarray}
\tilde\epsilon_{0}(0) < \tilde\epsilon_{0}(E_{l}) < \tilde\epsilon_{1}(0) < \tilde\epsilon_{1}(E_{l}) < ... & \quad\mbox{if}\quad
E_{l} > 0\quad,
\label{eq:gonze:plus}
\\[2ex]
\tilde\epsilon_{0}(E_{l}) < \tilde\epsilon_{0}(0) < \tilde\epsilon_{1}(E_{l}) < \tilde\epsilon_{1}(0) < ... & \quad\mbox{if}\quad
E_{l} < 0 \quad.
\label{eq:gonze:minus}
\end{eqnarray}
Now the actual KB-pseudopotential gives $E_l \equiv E^{\rm KB}_{l}$.
The reference eigenvalue $\epsilon^{\rm ps}_{l}$ which it reproduces
by construction corresponds to one of the levels 
$\{\tilde\epsilon_{0,1,\ldots}(E_{l}=E^{\rm KB}_{l})\}$.
If it corresponds to the ground state level, i.e.
$\epsilon^{\rm ps}_{l} \equiv \tilde\epsilon_{0}(E_{l}=E^{\rm KB}_{l})$,
obviously no lower (ghost) state exists. Whether this holds true 
is decided from (\ref{eq:gonze:plus}) or (\ref{eq:gonze:minus})
with the help of the eigenvalues for the local potential:
There is no ghost level beneath the reference level $\epsilon^{\rm ps}_{l}$
for
\begin{eqnarray}
E^{\rm KB}_{l} > 0 & \quad\mbox{\parbox[t]{.7\linewidth}{
if the reference level lies below the first excited level of the local potential, 
$\tilde\epsilon_{0}(0) < \epsilon^{\rm ps}_{l} < \tilde\epsilon_{1}(0)$ in Eq.~(\ref{eq:gonze:plus}),}}
\label{th:gonze:plus}
\\
E^{\rm KB}_{l} < 0 & \quad\mbox{\parbox[t]{.7\linewidth}{if the reference level lies 
below the ground state level of the local potential, 
$\epsilon^{\rm ps}_{l} < \tilde\epsilon_{0}(0)$
Eq.~(\ref{eq:gonze:minus}).}}
\label{th:gonze:minus}
\end{eqnarray}
The program {\sf pswatch} reports ghost states based on this criterion. 

As a practical rule ghost-free KB-pseudopotentials may be obtained
with the $d$-component as the local potential except for ``two-shell'' 
situations like the transition metal elements where typically the 
$s$-component is used as the local potential. To eliminate a ghost state 
occurring for some $l$ one may 
(i) change $l_{\rm loc}$, i.e. use a different component of the semilocal 
pseudopotential as the local potential,
(ii) adjust the core cutoff radii $r^{c}_{l}$ of the offending component 
$V^{\rm ps}_{l}(r)$ or the local potential $V^{\rm ps}_{l_{\rm loc}}(r)$, 
respectively.
In doing so the basic transferability ought to be maintained
(see Sec.~\ref{sec:transferability}). Also, it is expedient to
take $l_{\rm loc} = l_{\rm max}$ in order to keep the number of
the (computationally intensive) projections implied by Eq. (\ref{eq:kb-potential-r})
small.
Modifying the above entities seeks to adjust $E^{\rm KB}_{l}$, i.e. the 
strength of the KB-pseudopotential relative to the local potential such
that either (\ref{th:gonze:plus}) or (\ref{th:gonze:minus}) is satisfied. 
Note that a repulsive enough local potential (achieved e.g. by increasing
$r^{c}_{l_{\rm loc}}$ or taking $l_{\rm loc} = 0$) rules out any ghost 
state: Such a potential eventually leads to a negative KB-energy (\ref{eq:ekb}) 
as $\delta V^{\rm sl}_{l}(r) = V^{\rm ps}_{l}(r) - V^{\rm loc}(r) < 0$.
Also, $\tilde\epsilon_{0}(0) \rightarrow 0$ since such a local potential
gives at most a weakly bound ground state. As $E^{\rm KB}_{l} < 0$ one
has $\epsilon^{\rm ps}_{l} < \tilde\epsilon_{0}(0) \approx 0$ by virtue
of (\ref{eq:gonze:minus}), ruling out any ghost level.
\begin{table}
\caption{Analysis of ghost states for the Kleinman-Bylander pseudopotential
of Cu employing Eq.~\protect{(\ref{th:gonze:plus}) and (\ref{th:gonze:minus})}.
The eigenvalus for the local potential are underlined.
The semilocal pseudopotentials were constructed for the
$(3d^{10}\,4s^2\,4p^0)$-configuration, using the Troullier-Martins scheme
with cutoff radii $r^{c}_{4s}=r^{3d}_{l=2}\simeq 2.1 $~bohr and $r^{c}_{4p}\simeq 2.3$~bohr.
Exchange-correlation was treated within the LDA.}
\label{tab:kb:cu}
\vskip .5ex
\begin{tabular*}{\textwidth}{ccc@{\hspace{1em}}ccccc}
\hline
$l_{\rm loc}$
      & $l$
            & $E^{\rm KB}_{l}$ (eV)
                  & \multicolumn{4}{c}{bound state energies (eV)}
                        & ghost?\\
\hline
$2\,(3d)$
      & $0\,(4s)$
            & 319.67
      & \underline{$-219.06$}
            & $-72.98$
                  & \underline{$-15.95$}
                        & $-4.86$
                              & yes\\
      &&& $\tilde\epsilon_0(0)$
            & $\tilde\epsilon_0(E^{\rm KB}_{l})$
                  & $\tilde\epsilon_1(0)$
                        & $\tilde\epsilon_1(E^{\rm KB}_{l}) \equiv \epsilon^{\rm ps}_{4s}$\\
\cline{2-8}
      & $1\,\,(4p)$
            &  223.44
      & \underline{$-108.75$}
            & \multicolumn{2}{c}{$-0.77$}
                  & \underline{$-0.46$}
                        & no\\
      &&& $\tilde\epsilon_0(0)$
            & \multicolumn{2}{c}{$\tilde\epsilon_0(E^{\rm KB}_{l}) \equiv \epsilon^{\rm ps}_{4p}$}
                  & $\tilde\epsilon_1(0)$\\
\hline
$0\,(4s)$
      & $2\,(3d)$ 
            & $-273.80$
      & $-5.33$
            &&&& no\\
      &&& $\tilde\epsilon_0(E^{\rm KB}_{l}) \equiv \epsilon^{\rm ps}_{3d}$\\
\cline{2-8}
      & $1\,\,(4p)$
            &  51.64
      & \underline{$-0.92$} 
            & \multicolumn{2}{c}{$-0.77$}
                  && no\\
      &&& $\tilde\epsilon_0(0)$
            & \multicolumn{2}{c}{$\tilde\epsilon_0(E^{\rm KB}_{l}) \equiv \epsilon^{\rm ps}_{4p}$}\\
\hline
~\\
\end{tabular*}
\end{table}

Table \ref{tab:kb:cu} illustrates the foregoing for a copper pseudopotential
(see also Ref.~\cite{gon91a} and the examples in this package's distribution): 
the $3,4,5d$-component taken as local potential ($V^{\rm loc}=V^{\rm ps}_{l_{\rm max}}$) 
acts as a strongly attractive potential on the $s$-states. Accordingly the
eigenvalue sequence starts at quite low an energy. In the case of 
copper the reference energy of the $4s$-pseudopotential lies already 
above the first excited level for the local potential, 
$\tilde\epsilon_{1}(0) < \epsilon_{4s}$. By Eqs.~(\ref{eq:gonze:minus}) 
or (\ref{eq:gonze:plus}) the Cu $4s$-level corresponds to an excited 
$s$-like level of the KB-pseudopotential, and so there is a ghost state 
beneath the Cu $4s$-state. By contrast, taking the $s$-component as the 
local potential results in a ghost-free KB-pseudopotential.

\section{Description of the package}
\label{cha:package}

The package {\sf fhi98PP} provides as main utilities the UNIX C-shell scripts
{\sf psgen} and {\sf pswatch}.
The script {\sf psgen} and the related FORTRAN program {\sf fhipp} solve the
all-electron atom and generate the ionic pseudopotentials (see Secs. ~\ref{sec:atomic}
-- \ref{sec:unscreen}).
The script {\sf pswatch} and the related FORTRAN program {\sf pslp} solve the
pseudo-atom for a given pseudopotential, and carry out tests to monitor the 
pseudopotential's transferability in its semilocal and fully separable form
(see Secs. ~\ref{sec:transferability} -- \ref{sec:kb}).
The shell scripts serve as command line interfaces for running the FORTRAN
programs. For visual inspection of results, they group output into thematic 
graphic files to be viewed with the public domain, X-Windows based 
plotting tools XVGR or XMGR~\cite{tur91a}. 
The scripts include command summaries (execute {\sf psgen -h} or 
{\sf pswatch -h}). 
The Tables~\ref{tab:input:psgen} -- \ref{tab:input:sample} explain 
the required and the optional input of {\sf psgen} and {\sf pswatch}.
The Tables~\ref{tab:files:psgen} -- \ref{tab:format:psp}
summarize the meaning of the input and output files, and 
display their dependencies on the FORTRAN routines.
All input and output files reside in the chosen working directory.
The name of the input files can be chosen freely. For the output files,
the shell scripts compose filenames from a freely chosen mnemonic identifier 
and preset functional suffixes. For example, ``{\sf al.cpi}'' shall 
denote an output file which tabulates an aluminum pseudopotential.

The supplement ``Test run protocol'' at the end of this paper provides a 
prototype session protocol for generating pseudopotentials, for aluminum 
as an example.
A database of input files is distributed along with the package (see Sec.~\ref{sec:org}).
Below we outline the structure of the shell scripts and the related FORTRAN
programs, and then discuss the various input variables.

\subsection{
The tool {\sf psgen} for solving the all-electron atom and generating 
pseudopotentials}
\label{ssec:describe:psgen}

The script {\sf psgen} reads from the commandline the identifier string
{\it name}, and prinicpal input file {\sf $<${\it input data}$>$} containing
the atomic configuration and run parameters as explained below.
\begin{description}
\item
{\sf psgen -o {\it name} $<${\it input data}$>$}, the standard mode, performs the all-electron
calculation and constructs the pseudopotentials. Chief output is the run protocol 
{\sf {\it name}.dat} and the pseudopotential data file {\sf {\it name}.cpi}.
\item 
{\sf psgen -o {\it name} -fc {\it name}.fc $<${\it input data}$>$}, the frozen-core mode, 
performs
the all-electron calculation within the frozen-core approximation. There only 
the valence orbitals are determined self-consistently while the core density 
is held fixed as given by the file {\it name}.{\sf fc} (see below). This is
meant to generate a certain pseudopotential component from an excited
configuration with the unrelaxed core density of the ground state, and also to
establish benchmark values, e.g. for excitation energies, in transferability tests.
\item 
{\sf psgen -ao -o {\it name} $<${\it input data}$>$}, the atom-only mode, carries out just 
the all-electron calculation, as needed for instance to determine excitation 
energies.
\end{description}
The option {\sf -e} invokes an editor for the input file {\sf $<${\it input data}$>$}
The command {\sf psgen -h} lists all command options.

The prinicpal input file ({\sf $<${\it input data}$>$}, e.g. called {\sf al.ini}) specifies 
the electronic configuration of the atom, the partitioning of core and valence states, 
as well as parameters for handling exchange-correlation and constructing the 
pseudopotentials. Below we discuss the main variables, the Tables~\ref{tab:input:psgen} 
and \ref{tab:control:parameters} give an overview. Table~\ref{tab:input:sample} 
describes a sample file.

The variable $z$ stands for the atomic number. Non-integer values to realize
artificial atoms are allowed.

The variables $nc$ and $nv$ specify the number of core states and the number
of valence states, respectively. Note that the norm-conservation constraints 
can be met exactly for one and only one valence state per angular momentum 
channel (see Sec.~\ref{sec:screened}).

The variable $iexc$ specifies the exchange-correlation scheme used in solving
the all-electron atom and in the unscreening of the pseudopotentials.
Table~\ref{tab:input:sample} lists the implemented parameterizations of the
local-density approximation and the generalized-gradient approximations
(see Sec.~\ref{sec:atomic}). On the commandline $iexc$ is set by 
{\sf psgen -xc $iexc$ ...}\,. 

The variable $rnlc$ controls the handling of core-valence
exchange-correlation in the unscreening of the pseudopotential and the
construction of a partial core density (see~Sec. \ref{sec:unscreen}).
Setting $rnlc = 0$ corresponds to the ordinary linear treatment, a partial
core density is not computed. Taking $rnlc > 0$ implies the explicit
treatment of core-valence XC with the help of a partial core density.
The value of $rnlc$ specifies the cutoff radius (in bohr) beyond which the
partial core density is identical with the actual true core electron density.
The respective command is {\sf psgen -rc 1.3 ...}, e.g., setting
$rnlc = 1.3$~bohr. The partial core density is appended to the file
{\sf {\it name}.cpi}.
If the all-electron calculation is done in the frozen-core mode, the partial
core density is constructed from the input core electron density.

The variables $n(i)$, $l(i)$, and $f(i)$ denote the $i$-th state's principal
quantum number, angular momentum quantum number, and occupation number,
respectively.
The $i = 1, ..., nc$ core states need to be listed before the
$i =  nc+1, ..., nc+nv$ valence states. The principal quantum numbers should
appear in ascending order, $n(i) \geq n(i-1)$. Unoccupied valence states 
which are not listed are assumed to be ``unbound'', the respective pseudopotential 
is then calculated by the Hamann procedure (see Sec.~\ref{sec:screened}). 
Listing such states with $f(i) = 0$ instructs the program to determine 
the respective state. Note that negatively charged ions, 
$\sum_{i=1}^{nc+nv} f(i) > z$, are unstable within LDA or GGA, and may 
cause the calculation to fail.

The variable $lmax$ specifies the highest angular momentum channel up to
which the program constructs pseudopotential components. The variable
$s\_pp\_def$ allows to select either the Hamann scheme
$s\_pp\_def = \mbox{\sf h}$ or the Troullier-Martins scheme
$s\_pp\_def = \mbox{\sf t}$ for constructing the pseudopotentials.
Appropriate default cutoff radii are supplied by the program.
{\sf psgen -T $s\_pp\_type$ ...} is the related script command.

Any further input is optional and serves to override the default settings in
the angular momentum channel $lt$ for the cutoff radius $rct$ (in bohr), the
reference energy $ect$ (in eV), and the pseudopotential scheme  $s\_pp\_type$.
This feature is mainly for finetuning the cutoff radii when a ghost state
of the Kleinman-Bylander form has to be eliminated (see Sec.~\ref{sec:kb}). If
the cutoff radii are enlarged in order to achieve softer pseudopotentials, the
transferability of the resulting pseudopotential should be carefully verified.
On the commandline, the options {\sf -rs, -rp, -rd, -rf} may be used to set
the cutoff radii of the $s,p,d,f$-components of the pseudopotentials. For
instance {\sf psgen -rp 1.5 \dots} sets the $p$-cutoff radius to $1.5$~bohr.

By the commandline option {\sf -r} one instructs {\sf psgen} to perform
a non-relativistic all-electron calculation rather than a scalar-relativistic
one (default). The corresponding variable $tnrl$ is then set to true.

By the commandline option {\sf -g} all FORTRAN input and output files are
saved, rather than being removed upon exiting {\sf psgen}. 

The script {\sf psgen} first generates appropriate input files ({\sf fort.20} and 
{\sf fort.22}) used by the FORTRAN routine {\sf fhipp} which it invokes next. 
If the frozen-core mode is applied, the core density {\it name}.{\sf fc} is copied 
to {\sf fort.18} (see also below). 
After completion of {\sf fhipp} the script writes a descriptive
protocol to the file {\it name}.{\sf dat}. It then generates the file 
{\it name}.{\sf cpi} 
which contains the components of the ionic pseudotential, the pseudo wavefunctions, 
and, optionally, any partial core density on a logarithmic radial grid, adapted
to the format used in the {\sf fhi96md} electronic-structure and molecular-dynamics 
package~\cite{boc97a}. If
different reference configurations are used in constructing the pseudopotential
components, a customized file {\it name}.{\sf cpi} can be created from 
the corresponding FORTRAN output files {\sf fort.40}, etc., described in 
Table~\ref{tab:format:psp}.
The output file {\it name}.{\sf aep} contains the all-electron full potentials, and 
will be used by the utility {\sf pswatch} to evaluate the all-electron logarithmic 
derivatives. The file {\it name}.{\sf fc} lists the full core electron density, and
may be used as input to {\sf psgen} in a subsequent frozen-core calculation. 
Moreover the script produces various graphical files for visual inspection of, e.g., 
the ionic pseudopotential (file {\sf xv.{\it name}.pspot\_i}) or the all-electron
wavefunctions (file {\sf xv.{\it name}.ae\_wfct}).

In the pertinent FORTRAN program {\sf fhipp} the main routine {\sf fhipp} initially 
reads the run attributes and atomic configuration from the unformatted files 
{\sf fort.20} and {\sf fort.22}. A logarithmic radial mesh is set up,
$\lbrace r_{m} = a^{m-1}_{\rm mesh} r_{\rm min}(Z) \quad|\quad m = 1,2,...,m_{\rm max}\rbrace$,
with default parameters as set in the FORTRAN header file {\sf default.h}. All
numerical integrations are done on this mesh.
Any core electron density is read in from {\sf fort.18}, and must be tabulated 
on the current radial mesh.
The subroutine {\sf sratom\_n} self-consistently solves for the all-electron 
atomic eigenstates and the full potential. The radial Schr\"odinger equations
are integrated by a predictor-corrector scheme in the subroutine {\sf dftseq}.
The self-consistency cycle stops once all eigenvalues are converged.
The program then writes a run protocol to 
{\sf fort.23}, the all-electron wavefunctions to {\sf fort.38}, the full 
core density to {\sf fort.19}, and the full all-electron potential to 
{\sf fort.37}. If the atom-only mode is applied,
the program exits at this point. Otherwise the main routine next 
determines the default parameters for constructing the pseudopotentials,
and checks for any optional input in {\sf fort.22} to override the
default settings.
The subroutine {\sf ncpp} then determines the pseudo wavefunctions and the
screened pseudopotentials. The program proceeds with the unscreening of the
pseudopotentials. Optionally the subroutine 
{\sf dnlcc} constructs an appropriate partial core density, either from 
the self-consistent full core density or from the user provided frozen-core 
density. Finally a protocol of the pseudopotential construction 
is appended to {\sf fort.23}. The ionic pseudopotential and the pseudo 
wavefunction of each angular momentum channel $l$ are written to 
{\sf fort.4[$l=0,1,...$]}. The components of the screened pseudopotential
are written to {\sf fort.45} ($l=0$), {\sf fort.46} ($l=1$), etc.\,. 
The partial core density is output to {\sf fort.27}. All quantities are
tabulated on the dense logarithmic mesh that was employed in computing them, 
except for the all-electron wavefunctions which are output on a sparser 
mesh.

\subsection{The tool {\sf pswatch} for solving the pseudo atom and checking 
pseudopotentials}
\label{ssec:describe:pswatch}

The script {\sf pswatch} is normally run after {\sf psgen}. It reads from the 
command line the identifying string {\it name}, and the name of the principal
input file {\sf $<${\it input data}$>$} for the atomic valence configuration.  
The latter has the same layout as when generating the pseudopotentials 
(see Sec. \ref{ssec:describe:psgen}).  Furthermore {\sf pswatch} assumes the 
presence of the files {\it name}.{\sf cpi} (ionic pseudopotentials) and
{\it name}.{\sf aep} (all-electron potential). Both are routinely
provided by a run of {\sf psgen} as described above. The same radial
grid must be used in these files. Also, the all-electron potential
should be consistent with the present atomic configuration.
Several run modes are implemented:
\begin{description}
\item
{\sf pswatch -i {\it name} $<${\it input data}$>$}, the standard mode, performs first
a self-consistent calculation of the pseudo-atom, then computes the 
logarithmic derivatives for the all-electron potential and the screened
pseudopotential, and carries out the ghost state analysis for the
fully separable pseudopotential.
\item
{\sf pswatch -ao -i {\it name} $<${\it input data}$>$}, the atom-only mode, just solves
the pseudo-atom, e.g., in order to determine excitation energies.
\item
{\sf pswatch -kb -i {\it name} $<${\it input data}$>$}, the semilocal-only mode, skips
the ghost state analysis.
\end{description}
The option {\sf -e} invokes an editor for the input file {\sf $<${\it input data}$>$}.
The command {\sf pswatch -h} gives a list of all options. 

Several parameters that control the evaluation of the logarithmic 
derivatives and the handling of the fully separable pseudopotential 
can be specified on the command line. The script {\sf pswatch} collects 
them in an auxiliary input file ({\sf fort.21}). Below we introduce the
main parameters, the Tables~\ref{tab:input:psgen} and 
\ref{tab:control:parameters} list them in full. Sample input files are 
given in Table~\ref{tab:input:sample}.

The variable $lmax$ specifies the angular momentum channel up to which 
pseudopotential components are present. For $l \leq lmax$ the program 
uses the respective components to compute the logarithmic derivatives. 
For $l > lmax$ it uses the local potential instead.

The variable $lloc$ denotes the angular momentum channel whose 
pseudopotential component should be taken as the local potential, with
$lloc \leq lmax$. For instance the $s$-component is chosen by the 
{\sf pswatch -l 0 ...} command.

The variable $rlgd$ gives the diagnostic radius (in bohr) at which the 
logarithmic derivatives are evaluated. If $rlgd = 0$ the program assigns 
as a default value the covalent radius for the atom at hand. 
Note that the diagnostic radius should be larger than the cutoff radii used 
in constructing the pseudopotential. The according command is, e.g., 
{\sf pswatch -rl 2.5 \dots}, setting $rlgd  = 2.5$~bohr.

The commandline option {\sf pswatch -cpo $<${\it outfile}$>$ ...} causes
the program to use for the analysis of the Kleinman-Bylander form 
the self-consistently calculated pseudo wavefunctions rather than those 
provided by the file {\it name.{\sf cpi}} (default). The calculated 
wavefunctions will be written to a file {\it outfile} analogous 
to {\it name.{\sf cpi}}.  This feature may be used, e.g., when just 
the ionic pseudopotential but not the respective pseudo wavefunctions 
are known. The corresponding variable {\em tiwf} true is then set true.

The commandline option {\sf -r} instructs {\sf pswatch} to calculate
the logarithmic derivatives for the all-electron potential
non-relativistically rather than scalar-relativistically (default). 
The corresponding variable $tnrl$ is then set true.

By the commandline option {\sf -g} all FORTRAN input and output files
are saved, rather than being removed upon exiting {\sf psgen}.

The script sets up appropriate input for the FORTRAN program {\sf pslp} which
is invoked next.
A self-evident protocol is printed to the terminal and also written to the file 
{\it name}.{\sf test}. Upon completion, {\sf pswatch} assembles the file
{\sf xv}.{\it name}.{\sf lder} for the visual inspection of the logarithmic 
derivatives of the all-electron potential and pseudopotentials.

In the FORTRAN program {\sf pslp} the main routine {\sf pslp} reads the
run attributes and atomic valence configuration from the unformatted
files {\sf fort.21} and {\sf fort.22}, the latter having the same layout 
as for the {\sf fhipp} program. Next it reads from {\sf fort.31} the
(logarithmic) radial mesh, the pseudo wavefunctions to be used in the Kleinman-Bylander 
construction, the components of the ionic pseudopotential,
and, if present, the partial core density. 
The program writes a run protocol to standard output.
The subroutine {\sf psatom} first self-consistently solves for the
wavefunctions and the screening potential of the pseudo atom. The
self-consistency cycle stops once all eigenvalues have converged.
The program then writes 
the semilocal ionic pseudopotential and the computed pseudo
wavefunctions to {\sf fort.4[$l=0,1,...$]}, the related
screened pseudopotential to {\sf fort.45, fort.46, }etc., and
any partial core density to {\sf fort.27}. If the atom-only mode 
is applied, the program stops at this point. Otherwise the subroutine
{\sf ppcheck}
proceeds with the evaluation of logarithmic derivatives. This is done
for a user-specified or a preset diagnostic radius. An 
adaptive energy mesh is employed, where the upper and lower 
bounds may be adjusted in the FORTRAN source file {\sf ppcheck.f}.
As the first step, the all-electron potential is read in from the file 
{\sf fort.37}. It must be given on the same radial mesh as the pseudopotentials. 
Integrating the radial Schr\"odinger equation for each angular momentum channel $l$,
the all-electron logarithmic derivative is output to {\sf fort.5[$l=0,1,...$]}.
The corresponding logarithmic derivative of the self-consistently 
screened semilocal pseudopotential is tabulated in {\sf fort.6[$l=0,1,...$]}.
Next, {\sf ppcheck} 
transforms the pseudopotential into the fully separable form (see 
Sec.~\ref{sec:kb}).
By default the inputted pseudo wavefunctions (file {\sf {\it name}.cpi}) 
are used to set up the projector functions, optionally they are substituted 
by the pseudo wavefunctions obtained earlier in {\sf psatom}. 
The subroutine {\sf derlkb} then evaluates the logarithmic derivative
for the respective nonlocal radial Schr{\"o}dinger equation, which is 
output to {\sf fort.7[$l=0,1,...$]}. In the next step, {\sf klbyii} 
computes the bound state spectra for the semilocal and nonlocal 
pseudopotentials. Furthermore {\sf ppcheck} examines the fully separable 
potentials for ghost states. Finally the subroutine {\sf kinkon} 
evaluates the kinetic energy for each calculated pseudo wavefunction 
in momentum space (see Sec.~\ref{sec:transferability}).

\subsection{Set-up of the package}
\label{sec:org}

The package {\sf fhi98PP} is distributed as a tar-archive which can be unpacked 
by the UNIX command {\it tar}. The directory {\sf Dfhipp} forms the root of the 
package's directory tree.

In the directory {\sf Dplay} we supply sample input and output files for several
exemplary pseudopotentials (aluminum, sodium, arsenic, selenium, and nitrogen),
along with a detailed tutorial (as Postscript file). These examples should
illustrate on a rather ``hands-on'' level the procedure for constructing 
pseudopotentials.

The directory {\sf bin/Tools} contains the shell scripts to generate ({\sf psgen}) 
and validate ({\sf pswatch}) the pseudopotentials. Also provided are some self-explanatory 
shell scripts ({\sf psdata\_al}, {\sf pstrans}) to assist with the transferability tests 
discussed in Sec.~\ref{sec:transferability}. 
In the directory {\sf bin/Elements} we supply a database of inputfiles used in 
constructing the pseudopotentials.  These have the generic names 
{\it chemical\_symbol}{\sf :h.ini} 
for the Hamann, and {\it chemical\_symbol}{\sf :tm.ini} for the 
Troullier-Martins scheme. 
The directory {\sf bin/Xvgr} contains the header files for the graphics output 
produced by the shell scripts. 

The directory {\sf src} includes the sources and libraries of the program, as well
as the {\sf Makefile} for the compilation of the FORTRAN programs {\sf fhipp} and
{\sf pslp} that are invoked by the shell scripts. Both programs use linear algebra 
routines
which need to be linked either from the ESSL-library or from a subset of the public 
domain LAPACK-routines. The latter is included as a library (libFREE) in the directory 
{\sf lib}, and may be compiled either as specified in the {\sf Makefile}, or, 
alternatively, simply along with the other routines.

In order to run the package's tools {\sf psgen} or {\sf pswatch} one has to create 
the executables {\sf fhipp.x} and 
{\sf pslp.x}. To this end one first needs to properly set in the {\sf Makefile} the 
name of the FORTRAN compiler, the compiler and linker flags, and specify whether the 
ESSL or libFREE library should be invoked. Executing the UNIX command {\it make all}
then does the compilation. Secondly, one needs to specify in the scripts {\sf psgen} and
{\sf pswatch} the correct source paths for the FORTRAN executables and the graphics
header files. We suggest to install the shell scripts {\sf psgen} and {\sf pswatch} 
as personalized commands.
Note that for viewing the graphics output from {\sf psgen} and {\sf pswatch} 
one of the public domain packages XMGR or XVGR must be available as well.

\section{Test run}

In the test run one calculates a Hamann-type pseudopotential for aluminum. 
To this end one executes {\sf psgen al -o al:h} as command. This  will copy
the inputfile {\sf al:h.ini} from the database {\sf bin/Elements} into the 
present working 
directory,
run the program {\sf fhipp} and process its outputfiles into the various text and
graphics files as described above. Secondly, one evaluates the logarithmic derivatives
and performs the ghost state analysis. This is done by executing the command 
{\sf pswatch -i al:h al:h.ini}, which will produce output on the terminal as well
as to various text and graphics files described in Sec.~\ref{cha:package}. 
Samples of the principal output and a terminal session protocol are listed in 
the Section ``Test run protocol''.
A full set of the input and output files of the test run is given in the 
directory {\sf sample}. Further examples can be found in the directory {\sf Dplay}.

\section*{Acknowledgements}

We thank M. Bockstedte, E. Pehlke, and A.P. Seitsonen
for numerous discussions and their effort in testing
this program. We are grateful to D.R. Hamann, N. Troullier
and J.L. Martins for granting the use of their codes, which
served as a starting point for the present work.

\section*{Appendix}
\label{app:describe-modes}

This appendix lists total energy expressions computed by
the utilities {\sf psgen} and {\sf pswatch}. 

{\em All-electron atom} --
The total energy of the spherical atom for a given set of occupancies
$f_{nl}$ (corresponding to a ground- or excited state) is obtained 
from the radial wavefunctions $u_{nl}(r)$ and electron density 
$\dens(r)=\frac{1}{4\pi r^2}\sum_{nl} f_{nl} |u_{nl}(r)|^2$ as
layed out in Sec.~\ref{sec:atomic},
\begin{equation}
\label{eq:etot:scf}
E^{\rm tot-AE}[\dens] = T[\dens] + E^{\rm XC}[\dens] + E^{\rm H}[\dens] + \int -\frac{Z}{r}\dens(r)\, d^3r
\quad,
\end{equation}
with the Hartree energy $E^{\rm H}[\dens] = \frac{1}{2}\int V^{\rm H}[\dens;r] \dens(r) \, d^3r$ 
(\ref{eq:vh}), the exchange-correlation energy $E^{\rm XC}$ in LDA (\ref{eq:exc-lda}) or 
GGA (\ref{eq:exc-gga}), and the kinetic energy 
\begin{equation}
\label{eq:kinetic-energy}
T[\dens] = \sum_{nl} f_{nl} \epsilon_{nl} - \int V[\dens;r] \dens(r)\, d^3r \quad,
\end{equation}
stated in terms of the one-particle or orbital energy,
$\sum_{nl}f_{nl}\epsilon_{nl}$,
and the potential energy due to the effective potential $V[\dens;r]$ (\ref{eq:vae}).

{\em Frozen-core all-electron atom} --
When the atom is calculated in the frozen-core approximation only the valence states are
determined self-consistently. The core density is taken over from a previously solved
atomic reference state, in the following indicated by zero sub- and superscripts. The 
effective potential (\ref{eq:vae}) is obtained with the electron density
\begin{equation} 
\dens(r) = \dens^{\rm core}_{0}(r) + \dens^{\rm val} \quad\mbox{with}
\end{equation}
\begin{equation}
\label{eq:core-density}
\dens^{\rm core}_{0}(r) = \frac{1}{4\pi r^2}\sum_{nl}^{\rm core} f_{nl}
\left|u_{nl}^{(0)}(r)\right|^2\quad,\quad\mbox{and}\quad
\dens^{\rm val}(r) =  \frac{1}{4\pi r^2}\sum_{nl}^{\rm val} 
f_{nl}\left|u_{nl}(r)\right|^2\quad,
\end{equation}
and the kinetic energies of the core and valence electrons are given by
\begin{equation}
T^{\rm core}[\dens_{0}] =
\sum_{nl}^{\rm core} f_{nl} \epsilon_{nl}^{(0)} - \int V[\dens_{0};r] \dens^{\rm core}_{0}(r) d^3r
\quad,
\end{equation}
\begin{equation}
T^{\rm val}[\dens] = \sum_{nl}^{\rm val} f_{nl} \epsilon_{nl}  - \int V[\dens;r] \dens^{\rm val}(r) d^3r
\quad.
\end{equation}
The atomic total energy in the frozen-core approximation then reads
\begin{equation}
\label{eq:etot:fc}
E^{\rm tot-FC}[\dens_{0};\dens] = T^{\rm core}[\dens_{0}] + T^{\rm val}[\dens] 
+ E^{\rm XC}[\dens] + E^{\rm H}[\dens] + \int -\frac{Z}{r}\dens(r) d^3r	\quad.
\end{equation}
For the reference configuration, i.e. $\dens \equiv \dens_{0}$, the 
unconstrained and frozen-core total energies (\ref{eq:etot:scf}) and 
(\ref{eq:etot:fc}) take on the same value.

{\em Pseudo atom} --
For a given set of valence occupancies $f_{l}$ the pseudo valence states
are determined by self-consistently solving the radial Schr\"odinger 
equations associated with the semilocal pseudopotential,
\begin{equation}
\left[ -\frac{1}{2}\frac{d^2}{dr^2} + \frac{l(l+1)}{2r^2} 
+ V^{\rm HXC}(r) + V^{\rm ps}_{l}(r)
- \epsilon^{\rm ps}_l \right] u^{\rm ps}_{l}(r) = 0 \quad,
\end{equation}  
with the screening potential
\begin{equation}
V^{\rm HXC}(r)
= V^{\rm XC}[\dens^{\rm ps}+\tilde\dens^{\rm core}_0;r] + V^{\rm H}[\dens^{\rm ps};r]
\quad,
\end{equation}
and 
$\dens^{\rm ps}(r) = \frac{1}{4\pi r^2}\sum_{l} f_{l} \left|u^{\rm ps}_{l}(r)\right|^2$. The total energy of the pseudo atom is given by
\begin{equation}
E^{\rm tot-PS}[\dens^{\rm ps}] = 
T[\dens^{\rm ps}] + E^{\rm XC}[\dens^{\rm ps}+\tilde\dens^{\rm core}_{0}] + E^{\rm H}[\dens^{\rm ps}] + 
\sum_{l} f_{l} \int_{0}^{\infty} V^{\rm ps}_{l}(r) |u^{\rm ps}_{l}(r)|^2\, dr \quad,
\end{equation}
where the kinetic energy associated with the pseudo valence states is
\begin{equation}
T[\dens^{\rm ps}] = \sum_{l} f_{l} \left( \epsilon^{\rm ps}_l 
- \int_{0}^{\infty} \left\{ V^{\rm HXC}(r)
+ V^{\rm ps}_{l}(r) \right\} |u^{\rm ps}_{l}(r)|^2\, dr \right)
\quad.
\end{equation}
For the nonlinear core-valence exchange-correlation scheme, $E^{\rm XC}$ 
and $V^{\rm XC}$ depend on the particular partial core density 
$\tilde\dens^{\rm core}_{0}(r)$ (see Sec.~\ref{sec:unscreen}).


\clearpage

\begin{table}
\caption{Atomic configuration and pseudopotential parameters, given in the 
input file {\it name}.{\sf ini} of the scripts {\sf psgen} and {\sf pswatch} 
(file {\sf fort.22} of the FORTRAN routines).}
\label{tab:input:psgen}
\end{table}
\tablefirsthead{\hline variable    & type & values & \\ \hline}
\tablehead{\hline variable    & type & values & \\ \hline}
\begin{supertabular*}{\textwidth}{
	p{2cm}@{\extracolsep{\fill}}
	p{2cm}@{\extracolsep{\fill}}
	c@{\extracolsep{\fill}}
	p{9.5cm}}
\multicolumn{4}{l}{required input, {\sf psgen} and {\sf pswatch}}\\[-1ex]
$ z$	&    {real}		& $>\,0$ 	& atomic number (nuclear charge)\\
$ nc$	&    {integer}	& $\geq\,0$ & number of core states\\
$ nv$	&    {integer}	& $\geq\,0$	& number of valence states\\
$ rnlc$  &    {real}       & $=\,0$    & linearized core-valence exchange-correlation \\[-1ex]
		&			& $>\,0$	& nonlinear core-valence exchange-correlation,\\[-1.1ex]
            &                 &           & using a partial core inside the radius {\it rnlc} (in bohr)\\
$ iexc$  &    integer      &           & exchange$^{\rm x}$-correlation$^{\rm c}$ approximation\\[-1ex]
		&			& 1		& \parbox[t]{3em}{LDA} Wigner$^{\rm c}$ \cite{wig34a}\\[-1ex]
		&			& 2		& \parbox[t]{3em}{LDA} Hedin/Lundquist$^{\rm c}$ \cite{hed71a}\\[-1ex]
		&			& 3		& \parbox[t]{3em}{LDA} Perdew/Zunger$^{\rm c}$ '81 \cite{per81a}\\[-1ex]
		&			& 4		& \parbox[t]{3em}{GGA} Perdew/Wang$^{\rm xc}$ '91 \cite{per92b}\\[-1ex]
		&			& 5		& \parbox[t]{3em}{GGA} Becke$^{\rm x}$, Perdew$^{\rm c}$ '86 \cite{bec88a,per86a}\\[-1ex]
		&			& 6		& \parbox[t]{3em}{GGA} Perdew/Burke/Ernzerhof$^{\rm xc}$ \cite{per96b}\\[-1ex]
 		&			& 7		& \parbox[t]{3em}{LDA} Perdew/Wang$^{\rm c}$ '92 \cite{per92a} $+$
 relativistic\\[-1ex]
		&			&		& \makebox[3em]{}      correction$^{\rm x}$ of MacDonald/Vosko \cite{mac79a}\\[-1ex]
		&			& 8		& \parbox[t]{3em}{LDA} Perdew/Wang$^{\rm c}$ '92 \cite{per92a}\\[-1ex]
		&			& 9		& \parbox[t]{3em}{GGA} Becke$^{\rm x}$, Lee/Yang/Parr$^{\rm c}$ \cite{bec88a,lee88a}\\[-1ex]
		&			& 10		& \parbox[t]{3em}{GGA} Perdew/Wang$^{\rm x}$ '91, 
Lee/Yang/Parr$^{\rm c}$ \cite{per92b,lee88a}\\
$n(i)$ 	&    {integer}	& 	      & 
\parbox[t]{9cm}{
principal quantum number of state $ i$\\
states should be in ascending order w.r.t. to $ n(i)$
}\\
$ l(i)$	&    {integer}	&      	& angular momentum quantum number of state $i$\\
$ f(i)$	&    {real}	      &      	& occupancy number of level $i$\\
\hline
\multicolumn{4}{l}{required input, {\sf psgen} only}\\[-1ex] 
$ lmax$	&    {integer}	& 0 ... 4   & maximum $l=l_{\rm max}$ to generate pseudopotential for\\
$ s\_pp\_def$ &    {character}  & h		& Hamann pseudopotential scheme\\[-1ex]
		&			& t         & Troullier/Martins pseudopotential scheme\\
\hline
\multicolumn{4}{l}{optional input, {\sf psgen} only}\\[-1ex] 
$ lt$	&    {integer}	& 0 ... 4   &
for $ lt$-th pseudopotential component:\hspace{1ex}
use alternate cutoff radius and/or reference energy and/or scheme from present line \\
$rct$	&    {real}		& $=\,0$	& core cutoff radius, use default \\[-1ex]
		&			& $>\,0$ 	& use this value (in bohr) instead\\
$et$	&    {real}		& $=\,0$	& 
reference energy, use default: \hspace{+1ex} for an occupied state 
its eigenvalue, for an unoccupied state the highest eigenvalue of the occupied states
\\
		&			& $\neq\,0$ & use this value (in eV) instead \\ 	
$ s\_pp\_type$ &  {character} & -        & pseudopotential scheme, use $s\_pp\_def$\\[-1ex]
            &                 & h, t      & see required input\\
\hline
\end{supertabular*}

\begin{table}
\caption{Upper section: Control parameters for the FORTRAN routine {\sf fhipp} 
invoked by the script {\sf psgen}, and passed by the input file {\sf fort.20}.
Lower section: The same for the FORTRAN routine {\sf pslp} invoked by the
script {\sf pswatch}, and passed by the input file {\sf fort.21}. These files
are created by and depend on the arguments given to the shell scripts.}
\label{tab:control:parameters}
\begin{tabular*}{\textwidth}{
	p{1.5cm}@{\extracolsep{\fill}}
	p{2cm}@{\extracolsep{\fill}}
	p{2cm}@{\extracolsep{\fill}}
	p{9.5cm}}
\hline
variable       & type & values & \\
\hline
\multicolumn{4}{l}{file {\sf fort.20} of {\sf fhipp} (optional input)}\\
{\em tdopsp}   &   {logical}     & true    & compute pseudopotential (default)\\[-1ex]
               &                 & false   & atom-only mode:\hspace{1ex} 
stop after solving all-electron
atom\\
{\em tnrl}     &   {logical}     & true    & non-relativistic all-electron atom\\[-1ex]
                &                & false   & dto. scalar-relativistic (default)\\
\hline
\multicolumn{4}{l}{file {\sf fort.21} of {\sf pslp} (required input)}\\
{\em lloc}     &    {integer}    & $0 \ldots l_{\rm max}$ & use $lloc$-th pseudopotential component 
as the local potential\\
{\em lbeg}    &    {integer}    & $0 \ldots l_{\rm max}$  & 
lowest angular momentum channel to calculate logarithmic derivatives for\\
{\em lend}    &    {integer}    & $0 \ldots l_{\rm max}$  &
highest angular momentum channel to calculate logarithmic derivatives for\\
{\em lmax}    &    {integer}    & $l_{\rm max}\geq 0$  & angular momentum number up to which 
pseudopotential components are present for, apply the local potential in any higher channel\\
{\em rlgd}     & real            & $\geq\,0$ & 
diagnostic radius where the logarithmic derivatives are computed, ought
be outside the core region\\
{\em tlgd}     &    {logical}    & true      & evaluate logarithmic derivatives (default)\\[-1ex]
               &                 & false     & dto. skipped\\
{\em tkb}      &    {logical}    & true      & analyze Kleinman-Bylander potentials (default)\\[-1ex]
               &                 & false     & dto. skipped\\
{\em tiwf}     &    {logical}    & true      & use input pseudo wavefunctions for the
construction of Kleinman-Bylander potentials (default)\\[-1ex]
              &                 & false     & dto. but use calculated waves \\
{\em tnrl}     &    {logical}    & true      & assume non-relativistic all-electron calculation in the calculation of the logarithmic derivatives\\
              &                 & false     & dto. scalar-relativistic (default)\\
\hline
\end{tabular*}
\end{table}

\newcommand{\sh}{\space\space}
\begin{table}
\caption{Sample input files to construct and validate an aluminum pseudopotential. 
}
\label{tab:input:sample}
\begin{tabular}{p{\textwidth}}
\hline
\parbox[t]{\textwidth}{
Atomic configuration and pseudopotential parameters \\
(file {\sf al.ini} of shell scripts, file {\sf fort.22} of FORTRAN routines) 
}
\\[+2ex]
\begin{minipage}[t]{6cm}
{\tt
13.00\sh  3\sh  2\sh  8\sh  0.0\\
1\sh     0\sh  2.00\\
2\sh     0\sh  2.00\\
2\sh     1\sh  6.00\\
3\sh     0\sh  2.00\\
3\sh     1\sh  1.00\\
2\sh h\\
0\sh 1.25\sh 0.00\sh h\\
1\sh 1.40\sh 0.00\sh h\\
}
\end{minipage}
\begin{minipage}[t]{6cm}
{\tt
z \sh nc \sh nv \sh iexc \sh rnlc\\
n(i) \sh l(i) \sh f(i)\\
.\\
.\\
.\\
.\\
lmax \sh s\_pp\_def\\
lt \sh rct \sh et \sh s\_pp\_type\\
.\\
}
\end{minipage}
\\[1ex]
\hline

control parameters of {\sf fhipp} (file {\sf fort.20} created by {\sf psgen})\\
\begin{minipage}[t]{6cm}
{\tt
.t.\sh .t.}
\end{minipage}
\begin{minipage}[t]{6cm}
{\tt
tdopsp\sh tnrl
}
\end{minipage}
\\[2ex]
\hline
control parameters of {\sf pslp} (file {\sf fort.21} created by {\sf pswatch})\\
\begin{minipage}[t]{6cm}
{\tt
2\sh 0\sh 2\sh 2\sh 0.0 \\ 
.t.\sh .t.\sh .t.\sh .f.
}
\end{minipage}
\begin{minipage}[t]{6cm}
{\tt
lloc\sh lbeg\sh lend\sh lmax\sh rld\\
tlgd\sh tkb\sh tiwf\sh tnrl\\
}
\end{minipage}
\\
\hline
\end{tabular}
\end{table}

\begin{table}[H]
\caption{File dependencies for the {\sf psgen} program. The upper part
refers to input files , the lower one to output files.}
\label{tab:files:psgen}
\begin{tabular*}{\textwidth}{
	p{6.0cm}@{\extracolsep{\fill}}
	p{4.0cm}@{\extracolsep{\fill}}
	p{2.5cm}@{\extracolsep{\fill}}
	p{2.8cm}}
\hline
&
{\sf psgen} files
&
\multicolumn{2}{l}{FORTRAN files and routines}
\\
\hline
\parbox[t]{2.5cm}{
command options,\\
atomic configuration
}
& \parbox[t]{3.5cm}{
	{\it input data}\\
	(e.g. {\sf {\it name}.ini})
}
&
\parbox[t]{2.5cm}{
{\sf fort.20}\\
{\sf fort.22}
}
&
\parbox[t]{2.5cm}{
{\sf fhipp}
}
\\

\parbox[t]{6.0cm}{
core density\\
(in frozen-core mode)
}
&
{\sf {\it name}.fc}
&
\parbox[t]{2.3cm}{
{\sf fort.18}
}
&
\parbox[t]{2.3cm}{
{\sf sratom\_n}
}
\\[3ex]
\hline

\parbox[t]{6.0cm}{
 run protocol
}
&
{\sf {\it name}.dat}
&
\parbox[t]{2.3cm}{
{\sf fort.23}
}
&
\parbox[t]{2.3cm}{
{\sf fhipp}\\
{\sf ncpp}
}
\\

\parbox[t]{6.0cm}{
full core density
}
&
{\sf {\it name}.fc}
&
{\sf fort.18}
&
{\sf sratom\_n}
\\

\parbox[t]{6.0cm}{
 full potential
}
& {\sf {\it name}.aep} &
 \parbox[t]{2.3cm}{
{\sf fort.37}
}
& {\sf sratom\_n}
\\

\parbox[t]{6.0cm}{
ionic pseudopotentials,\\
pseudo wavefunctions,\\
partial core density
}
&
\parbox[t]{2.3cm}{
{\sf {\it name}.cpi}
}
&
\parbox[t]{2.3cm}{
{\sf fort.40} etc.\\
{\sf fort.27} 
}
&
\parbox[t]{2.3cm}{
{\sf ncpp}\\
{\sf dnlcc}
}
\\

\parbox[t]{6.0cm}{
screened pseudopotentials 
}
&
{\sf xv.{\it name}.pspot\_s}
&
{\sf fort.45} etc.
&
{\sf ncpp}
\\

\parbox[t]{6.0cm}{
ionic pseudopotentials
}
&
{\sf xv.{\it name}.pspot\_i}
&
{\sf fort.40} etc.
&
{\sf ncpp}
\\

\parbox[t]{6.0cm}{
pseudo wavefunctions vs.\\
all-electron wavefunctions
}
&
{\sf xv.{\it name}.ps\_ae\_wfct}
&
{\sf fort.39}
&
{\sf ncpp}
\\

\parbox[t]{6.0cm}{
full core, partial core,\\
pseudo valence density
}
&
{\sf xv.{\it name}.density}
&
\parbox[t]{2.3cm}{
{\sf fort.19}\\
{\sf fort.27}\\
{\sf fort.25}
}
&
\parbox[t]{2.3cm}{
{\sf sratom\_n}\\
{\sf ncpp}\\
{\sf dnlcc}
}
\\ 

\parbox[t]{6.0cm}{
all-electron wavefunctions
}
&
{\sf xv.{\it name}.ae\_wfct}
&
{\sf fort.38}
&
{\sf sratom\_n}
\\
\hline
\end{tabular*}
\end{table}

\begin{table}[H]
\caption{File dependencies for the {\sf pswatch} program. The upper part
refers to input files, the lower one to output files.}
\label{tab:files:pswatch}
\begin{tabular*}{\textwidth}{
      p{6.0cm}@{\extracolsep{\fill}}
      p{4.0cm}@{\extracolsep{\fill}}
      p{2.5cm}@{\extracolsep{\fill}}
      p{2.8cm}}
\hline
&
{\sf pswatch} files
&
\multicolumn{2}{l}{FORTRAN files and routines}
\\
\hline
\parbox[t]{6.0cm}{
command options,\\
atomic configuration
}
& {\sf {\it name}.ini}
&
\parbox[t]{2.3cm}{
{\sf fort.21}\\
{\sf fort.22}
}
&
{\sf pslp}
\\

\parbox[t]{6.0cm}{
ionic pseudopotentials
}
&
{\sf {\it name}.cpi}
&
{\sf fort.31}
&
{\sf pslp}
\\

\parbox[t]{6.0cm}{
partial core density\\
(if present)
}
&
{\sf {\it name}.cpi}
&
{\sf fort.31}
&
{\sf pslp}
\\

\parbox[t]{6.0cm}{
all-electron potential
}
&
{\sf {\it name}.aep}
&
{\sf fort.37}
&
{\sf ppcheck}
\\

\hline
\parbox[t]{6.0cm}{
run protocol
}
&
\parbox[t]{2.3cm}{
{\it terminal},\\
{\sf {\it name}.test}
}
&
{\it std.out }
&
\parbox[t]{4.0cm}{
{\sf pslp}\\
{\sf ppcheck}\\
{\sf klbyii}\\
}
\\

\parbox[t]{6.0cm}{
ionic pseudopotentials,\\
calculated pseudo waves,\\
partial core density
}
&
{\it as specified}
&
{\sf fort.40} etc.
&
{\sf pslp}
\\

\parbox[t]{6.0cm}{
pseudo wavefunctions
}
&
---
&
{\sf fort.38}
&
{\sf psatom}
\\

\parbox[t]{6.0cm}{
logarithmic derivatives,\\
all-electron, semilocal,\\ 
fully separable case
}
&
{\sf xv.{\it name}.lder}
&
\parbox[t]{2.3cm}{
{\sf fort.50} etc.\\
{\sf fort.60} etc.\\
{\sf fort.70} etc.
}
&
\parbox[t]{2.3cm}{
{\sf ppcheck}\\
{\sf derlkb}
}
\\

\parbox[t]{6.0cm}{
ionic pseudopotentials
}
&
{\sf xv.{\it name}.pspot\_i}
&
{\sf fort.40} etc.
&
{\sf pslp}
\\

\parbox[t]{6.0cm}{
screened pseudopotentials
}
&
{\sf xv.{\it name}.pspot\_i}
&
{\sf fort.45} etc.
&
{\sf pslp}
\\[2ex]
\hline
\end{tabular*}
\end{table}

\begin{table}[H]
\caption{Format of the pseudopotential file {\it name}.{\sf cpi}
and the related FORTRAN output files {\sf fort.40}, etc., and {\sf fort.27}.}
\label{tab:format:psp}
\begin{tabular*}{\textwidth}{crc@{\extracolsep{\fill}}l}
\hline
line & col. & \multicolumn{2}{c}{description}\\
\hline
\multicolumn{4}{l}{file {\it name}.{\sf cpi} by script {\sf psgen}\,:}\\
1    & 1      & $Z^{\rm ion}$   & number of valence electrons\\[-1ex]
     & 2      & $l_{\rm max}+1$ & number of pseudopotential components\\
$2 \ldots 11$ &       &     & unused\\
12    & 1      & $m_{\rm max}$ & number of radial mesh points\\[-1ex]
     & 2      & $a_{\rm mesh}$ & mesh increment $r_{m+1}/r_{m}$\\
$12 \ldots m_{\rm max}+11$ & 1 & $m$  & radial mesh index\\[-1ex]
&              2 & $r_{m}$ & radial coordinate (in bohr)\\[-1ex]
&              3 & $u^{\rm ps}_{l}(\epsilon^{\rm ps}_{l};r_{m})$ & 
\parbox[t]{8cm}{
for $l=0$: radial pseudo wavefunction, \\
normalized as $\int_{0}^{\infty}|u^{\rm ps}_{l}(r)|^2 dr = 1$
}
\\
&              4 & $V^{\rm ps}_{l}(r_{m})$ & for $l=0$: ionic pseudopotential
(in hartree)\\
\multicolumn{4}{l}{-- for each $l=1,2, \ldots, l_{\rm max}\,$:\, put a block like for $l=0$}\\
\multicolumn{4}{l}{
-- if core-valence exchange-correlation applies, $r^{\rm nlc} > 0$:\,
append {\sf fort.27}}\\
\hline
\multicolumn{4}{l}{output {\sf fort.40}, etc. by FORTRAN program {\sf fhipp}\,:}\\
1 & 1 & ``\#'' & marker\\
  & 2 & $m_{\rm max}$ & number of radial mesh points\\[-1ex]
  & 3 & $a_{\rm mesh}$ & mesh increment $r_{m+1}/r_{m}$\\[-1ex]
  & 4 & $l$ & current angular momentum channel\\[-1ex]
  & 5 & $r^{\rm c}_{l}$ & used cutoff radius\\[-1ex]
  & 6 & $Z$ & atomic number (nuclear charge)\\
$2 \ldots m_{\rm max}+1$ & 1 & $m$ & radial mesh index\\
	& 2 & $r_m$ & radial coordinate (in bohr)\\
	& 3 & $u^{\rm ps}_{l}(\epsilon^{\rm ps}_{l};r_m)$  & radial pseudowavefunction\\
	& 4 & $V^{\rm ps}_{l}(r_m)$ & ionic pseudopotential\\ 
\hline
\multicolumn{4}{l}{output {\sf fort.27} by FORTRAN program {\sf fhipp}\,:}\\
$1 \ldots m_{\rm max}$ 
& 1 & $r_{m}$ & radial coordinate (in bohr)\\[-1ex]
& 2 & $\tilde\dens^{\rm core}(r_{m})$ &
\parbox[t]{8cm}{
partial core density, normalized as\\
$4\pi \int_{o}^{\infty} \tilde\dens^{\rm core}(r) r^2\,dr = \tilde N^{\rm core}$,
the circumstantial number of electrons in the partial core
}\\
& 3 & & dto. 1$^{\rm st}$ radial derivative\\[-1ex]
& 4 & & dto. 2$^{\rm nd}$ radial derivative\\
\hline
\end{tabular*}
\end{table}

\clearpage
\section*{Test run protocol}

\begin{verbatim}
COMMANDLINE> psgen -v -o al:h al.ini

13.00  3  2  8  0.0 : z  nc  nv iexc rncl
1  0  2.00          : n  l   f
2  0  2.00
2  1  6.00
3  0  2.00
3  1  1.00
2  h                : lmax  s_pp_def
0  1.25  0.00  h    : lt  rct  et  s_pp_type
1  1.40  0.00  h
<QUIT> <RETURN>

psgen - done: output
al.ini
al:h.aep
al:h.cpi
al:h.dat
al:h.fc
xv.al:h.ae_wfct
xv.al:h.density
xv.al:h.ps_ae_wfct
xv.al:h.pspot_i
xv.al:h.pspot_s
xv.al:h.unscreen

COMMANDLINE> more al:h.dat

fhi pseudopotential tool fhipp - version rev 1998

               chemical symbol  Al
                nuclear charge  13.00
                  total charge    .00
         number of core states   3
      number of valence states   2
    exchange-correlation model   8  LDA CA Perdew/Wang 1991       
      scalar-relativistic mode
        parameters radial mesh   493    1.024700   .480769E-03

       === all-electron atom ===

<        n     l      occupation  eigenvalue(eV)

<  1     1     0        2.0000        -1504.3103
<  2     2     0        2.0000         -107.5091
<  3     2     1        6.0000          -69.7240
<  4     3     0        2.0000           -7.8302
<  5     3     1        1.0000           -2.7840

                                  (Hartree a.u.)
                  total energy        -241.76605
                kinetic energy         241.94185
                orbital energy        -134.51716
                coulomb energy        -579.08935
                hartree energy         112.85767
   exchange-correlation energy         -17.47621
           xc potential energy         -23.08499
          number of iterations                29   convergence  0.0E+00
            integrated density          13.00000
 
 fhipp - all-electron atom done

    === HAMANN mode ===   h     

  l  n     radius:     node      peak       default core
x 0  3                 .800      2.023      1.214
x 1  3                 .800      2.582      1.549
x 2  3                 .000       .000      1.549

    === pseudo atom ===

  l  type  rcore       rmatch          eigenvalue(eV)      norm test   slope test
                                 all-electron     pseudo     1 =         1 =
  0  h     1.2418974   3.2163140  -7.8302025  -7.8302030   1.0000000   1.0000031
  1  h     1.3692182   4.4168841  -2.7839599  -2.7839599   1.0000000   1.0000024
  2  h     1.5468790   3.7233928  -2.7839599  -2.7839601   1.0000000   1.0000000
 
  --- linearized core-valence XC ---
        c-v equidensity radius           1.43769

                                  (Hartree a.u.)
                  total energy          -1.94588
                kinetic energy            .62119
              potential energy          -3.42557
                hartree energy           1.44497
                     xc energy           -.58647
    integrated valence density           3.00000
                  y range xvgr        -3   1   1
 
 fhipp - done for input @
 
@ 13.00  3  2  8 .00E+00 : z  nc  nv iexc rncl
@     1  0   2.00        : n  l   f
@     2  0   2.00
@     2  1   6.00
@     3  0   2.00
@     3  1   1.00
@ 2  h                   : lmax  s_pp_def
@ 0   1.25   .00E+00  h  : lt  rct  et  s_pp_type
@ 1   1.40   .00E+00  h

COMMANDLINE> pswatch -v -i al:h al.ini

2 0 2 2 0.0         : lloc lbeg lend lmax rlgd
.t. .t. .t. .f.     : tlgd tkb  tiwf tnrl
13.00  3  2  8  0.0 : z  nc  nv iexc rncl
1  0  2.00          : n  l   f
2  0  2.00
2  1  6.00
3  0  2.00
3  1  1.00
2  h                : lmax  s_pp_def
0  1.25  0.00  h    : lt  rct  et  s_pp_type
1  1.40  0.00  h
<QUIT> <RETURN>

fhi pseudopotential tool pslp - version rev 1998

               chemical symbol  Al
                nuclear charge  13.00
   number of valence electrons   3.00
      number of valence states   2
    exchange-correlation model   8  LDA CA Perdew/Wang 1991       
        parameters radial mesh   493    1.024700   .480769E-03
  input pseudopotentials up to   d

          === pseudo atom (Hartree a.u.) ===

<        n     l   occupation  eigenvalue(eV)  potential energy
<  1     1     0      2.0000       -7.8302        -1.21008
<  2     2     1      1.0000       -2.7840        -1.00541
 
                  total energy      -1.94588
                kinetic energy        .62119
  ionic pseudopotential energy      -3.42557
                hartree energy       1.44497
                     xc energy       -.58647
        local potential energy      -3.89662
           xc potential energy       -.76337
    integrated valence density       3.00000
          number of iterations            16   convergence  0.0E+00
                  y range xvgr        -3   1   1

 pslp- pseudoatom done - now testing

 --- assuming scalar-relativistic all-electron atom ---

 --- d component taken as local potential ---
 --- input wavefunctions used for kb potentials ---

 --- kb potentials: spectrum of bound states (eV) ---

            l          e0            e1            e2
semilocal   0       -7.8302        -.2102         .0000
nonlocal    0       -7.8302        -.2108         .0000
semilocal   1       -2.7838         .0000         .0000
nonlocal    1       -2.7838         .0000         .0000

 --- analysis of kb potentials: s waves  ---

 * no ghost (ekb > 0, eloc0 < eref < eloc1)
 
                     kb cosine         .3783
                     kb energy       38.3109 eV      ekb
   local potential groundstate      -23.2494 eV    eloc0
        dto. 1st excited state       -1.6612 eV    eloc1
              reference energy       -7.8302 eV     eref

 --- analysis of kb potentials: p waves  ---

 * no ghost (ekb > 0, eloc0 < eref < eloc1)
 
                     kb cosine         .3180
                     kb energy       18.3317 eV      ekb
   local potential groundstate       -6.8226 eV    eloc0
        dto. 1st excited state        -.0197 eV    eloc1
              reference energy       -2.7840 eV     eref

 --- logarithmic derivatives: at radius = 2.9893 ---

 --- nonlocal potentials ---
 --- all-electron potential ---
 --- semilocal potentials ---

 --- kinetic energy convergence in momentum space ---

     l  n  bracket   cutoff    norm   kinet. energy
            (eV)      (Ry)               (Hartree)
ck   0     1.0E+00      1    .981164    .147167E+00
ck   0     1.0E-01      9    .999384    .179742E+00
ck   0     1.0E-02     21    .999971    .182613E+00
ck   0     1.0E-03     30    .999998    .182930E+00
cx   0  1                   1.000000    .182961E+00

ck   1     1.0E+00      2    .959957    .221304E+00
ck   1     1.0E-01      3    .998345    .252144E+00
ck   1     1.0E-02      9    .999935    .254902E+00
ck   1     1.0E-03     16    .999996    .255235E+00
cx   1  2                   1.000000    .255268E+00

pswatch - done: output
al:h.test
xv.al:h.lder

COMMANDLINE>
\end{verbatim}

\vskip 5em
\hrulefill

\begin{thebibliography}{99}
\bibitem{com:relativity}{
The relativistic corrections to LDA exchange (Ref.~\cite{mac79a}) 
are significant mostly for the high densities close to the nucleus.
As pointed out in Ref.~\cite{bac82a} they affect mainly the deep 
$s$-states, yet have a minor effect on the valence levels. We usually 
omit these corrections, finding e.g. an upward shift of 0.01~eV for 
the $3s$-state of Zn, and of 0.04~eV for the $6s$-state of Hg compared 
to the nonrelativistic local-density description of exchange-correlation.
}
\bibitem{com:reference-level}{
The semilocal form of the pseudopotentials relies essentially on a 
single reference level for each $l$-component. A systematic generalization
to multiple reference levels using fully nonlocal pseudopotentials has been 
proposed by Vanderbilt and also by Bl\"ochl, see Ref.~\cite{blo89a}.
}
\bibitem{com:ekin-criterion}{
It is understood here that the pseudo wavefunctions' kinetic 
energy governs the plane wave convergence of the total energy. This may  
not hold true when a partial core density is present and is treated in 
reciprocal space, requiring a larger cutoff energy than implied by
Eq.~(\ref{eq:ekin-criterion}).
}
\bibitem{com:projectors}{
In Eqs. (\ref{eq:ukb}) and (\ref{eq:ekb}) we follow the notation of 
Ref.~\cite{gon91a}. Note however that equivalent definitions using
differently normalized projector functions are possible.
}
\end{thebibliography}
\end{document}